\documentclass[aps,pra,twocolumn,preprintnumbers,amsmath,amssymb,lineno,footinbib]{revtex4}
\usepackage{graphicx,epsfig}
\usepackage{bm}
\usepackage{dcolumn}

\usepackage{graphicx} 
\usepackage[colorlinks=true,linkcolor=blue,citecolor=blue,urlcolor=black]{hyperref}
\usepackage[english]{babel}

\usepackage{datetime}
\usepackage{verbatim}
\usepackage{amsmath}
\usepackage{amssymb}
\usepackage{siunitx} 
\usepackage{nccmath}
\usepackage{dcolumn}
\usepackage[switch,columnwise]{lineno}
\usepackage{xcolor}
\usepackage{color}
\usepackage{float}
\usepackage[utf8]{inputenc}
\usepackage{booktabs}

\begin{document}


\title{Direct visualization of superselective colloid-surface binding mediated by multivalent interactions}

\author{Christine Linne}
\affiliation{Soft Matter Physics, Huygens-Kamerlingh Onnes Laboratory, Leiden Institute of Physics,2300 RA Leiden, The Netherlands}
\affiliation{Department of Bionanoscience, TU Delft, 2629 HZ Delft, The Netherlands}
\author{Daniele Visco}
\affiliation{Department of Materials and Thomas Young Centre, Imperial College London, SW72AZ, London, UK}
\author{Stefano Angioletti-Uberti}
\affiliation{Department of Materials and Thomas Young Centre, Imperial College London, SW72AZ, London, UK}
\author{Liedewij Laan}
\email{Corresponding author: l.laan@tudelft.nl}
\affiliation{Department of Bionanoscience, TU Delft, 2629 HZ Delft, The Netherlands}
\author{Daniela J. Kraft}
\email{Corresponding author: kraft@physics.leidenuniv.nl}
\affiliation{Soft Matter Physics, Huygens-Kamerlingh Onnes Laboratory, Leiden Institute of Physics,2300 RA Leiden, The Netherlands}

\date{\today}
\begin{abstract}
Reliably distinguishing between cells based on minute differences in receptor density is crucial for cell-cell or virus-cell recognition, the initiation of signal transduction and selective targeting in directed drug delivery. Such sharp differentiation between different surfaces based on their receptor density can only be achieved by multivalent interactions.
 Several theoretical and experimental works have contributed to our understanding of this ``superselectivity'', however a versatile, controlled experimental model system that allows quantitative measurements on the ligand-receptor level is still missing. Here, we present a multivalent model system based on colloidal particles equipped with surface-mobile DNA linkers that can superselectively target a surface functionalized with the complementary mobile DNA-linkers. Using a combined approach of light microscopy and Foerster Resonance Energy Transfer (FRET), we can directly observe the binding and recruitment of the ligand-receptor pairs in the contact area. We find a non-linear transition in colloid-surface binding probability with increasing ligand or receptor concentration. In addition, we observe an increased sensitivity with weaker ligand-receptor interactions and we confirm that the time-scale of binding reversibility of individual linkers has a strong influence on superselectivity. These unprecedented insights on the ligand-receptor level provide new, dynamic information into the multivalent interaction between two fluidic membranes mediated by both mobile receptors and ligands and will enable future work on the role of spatial-temporal ligand-receptor dynamics on colloid-surface binding.
\end{abstract}

\maketitle
Processes at biological interfaces  are often governed by multivalent interactions. They play a key role in signal transduction, through inhibition and activation of signaling complexes, recognition and interactions between viruses and cells, as well as cell-cell adhesion \cite{Mammen1998,Huskens2006,Satav2015,Boudreau1999,Fasting2012}.
Multivalent bonds consist of a large number of weak bonds instead of a single strong one, which creates an interaction that is not only strong but also highly selective. The selectivity in multivalent systems goes beyond the correct recognition of a single ligand-receptor pair and allows ``superselective'' binding only to surfaces that exceed a critical receptor concentration. 
This allows for a sharp differentiaton of surfaces that consist of the same receptor type but vary in receptor density. Integrating this powerful feature into drug delivery systems would enable highly selective targeting of diseased cells \cite{Tian2020,Meng2020,Hauert2014}, for example, in cancer therapy \cite{Wang2020,Zhang2020_2,Zhang2019} where tumor cells over-express receptors on their surface \cite{Li2005,Akhtar2014}, or to target viral infections \cite{Koenig2021}.

For multivalent recognition and particle uptake in biological settings, as well as for applications such as directed drug delivery, the binding affinity to the target surfaces needs to be precisely tuned. To this end, the bond should be selective and strong yet weak enough to be reversible to, for example, facilitate endocytosis \cite{Doherty2009}. Specifically, theoretical studies have shown that the ligand density as well as interaction strength need to be adjusted with respect to the receptor density to increase the selective surface binding \cite{Martinez-Veracoechea2011,Duncan2015,Wang2012}. In addition, receptor mobility on the target surface - a key feature of membranes - leads to receptor clustering that can enhance the surface binding at low receptor concentrations \cite{Dubacheva2019,Lanfranco2019,DiIorio2020}.

These theoretical predictions have inspired the design of various experimental systems that can be used to investigate superselective surface binding. The ideal system for understanding superselectivity in biological context should mimic the lateral mobility of the ligands and receptors, and provide full control over their interaction strength and surface densities, as well as yield quantitative insights into the bond formation and dynamics. Experimentally, superselective surface binding has been demonstrated for systems that consisted of ligand-bearing polymers \cite{Dubacheva2014,Dubacheva2015,Dubacheva2019}, DNA-coated particles \cite{Lanfranco2019,Scheepers2020}, Influenza virus particles \cite{Overeem}, as well as small and giant unilamellar vesicles \cite{DiIorio2020}. These experiments confirmed theoretical predictions that a low binding affinity and high valency are crucial for obtaining superselectivity, and  furthermore showed that lateral mobility of receptors on the target surface can induce clustering of the ligands or receptors in the bond area, which enhances superselectivity \cite{Dubacheva2019}.

However, despite these intriguing observations, to date no system exists that captures the key features of biological interfaces, with both mobile ligands and receptors hosted by a lipid membrane, nor one that combines fluidic interfaces and the possibility for direct visualization of binding dynamics with a tunable interaction strength and ligand/receptor densities. This lack of a fully tunable model system prevents us from developing a comprehensive framework for multivalent interactions in biologically relevant settings. In particular, we expect that direct visualization of the spatial distribution of surface-mobile ligands and receptors will provide insights into their dynamics and impact on superselective surface binding.

Here, we introduce a colloid-based model system that allows direct investigation of the individual ligand-receptor interactions in a multivalent bond and their collective binding behaviour to a target surface. We achieve this using fluorescently labelled double-stranded DNA  with a single stranded overhang that can hybridize with the complementary sequence, anchored in a lipid membrane both on the colloids as well as the target surface. We observe that the colloid-target surface binding is mediated by the accumulation of receptors and ligands in the contact area. Interestingly, on the multivalent interaction timescale, individual ligand-receptor bonds dissolve and reform repeatedly. This dynamic reversibility confirms that the individual interactions are weak, in agreement with our observation of superselective binding at a critical ligand and receptor density. Our results motivate the development of novel theoretical models that link individual ligand-receptor dynamics to colloid-membrane binding and to reconcile effects taking place on the molecular scale with those on the micrometer scale.

\section*{Results}

\begin{figure}[tbhp]
\centering
\includegraphics[width=0.9\linewidth]{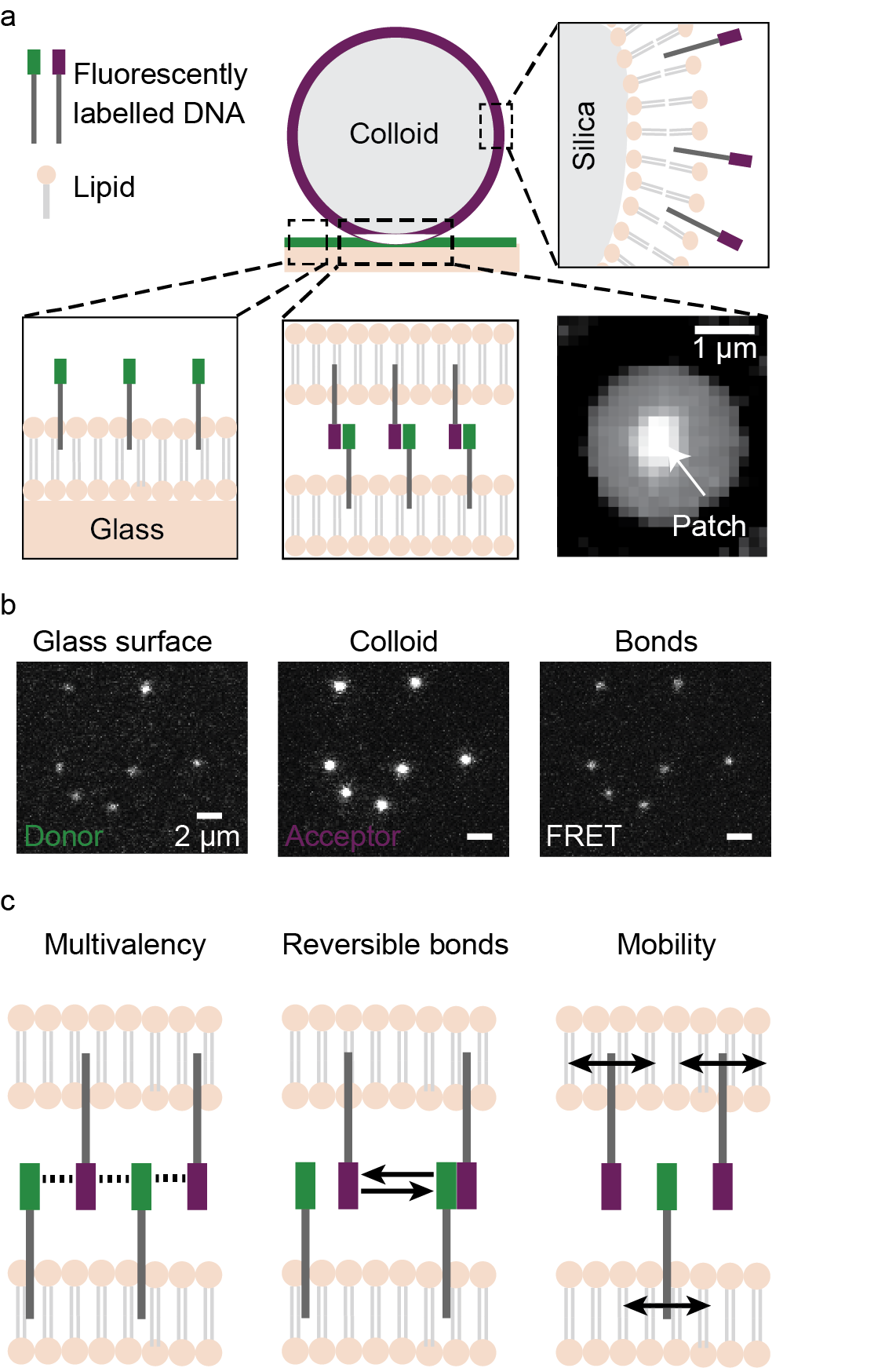}
\caption{\textbf{Experimental model system} (a) The two-dimensional experimental system consists of $2.12 \mu$m silica colloids  functionalized with double-stranded DNA stems with a single-stranded linker overhang as the ligand and receptor system. The DNA strands are anchored in a lipid membrane on both the colloid and flat glass surface which ensures their full mobility on the surface. (b) We separately detect the fluorescent signal of the receptor (left) and ligand (middle) DNA strands, as well as the FRET signal emitted by bound ligand-receptor pairs (right). The presence of a bright patch in the FRET channel indicates that a multivalent bond has been formed between the colloid and the surface. 
 (c) Our model system features surface mobility of both ligands and receptors and allows us to tune the number of ligands and receptors (valency) as well as the interaction strength to systemacially study superselective colloid-surface binding on the ligand-receptor level.}
\label{fig:Fig1}
\end{figure}

Our multivalent experimental model system consists of colloidal probe particles functionalized with ligands and a surface featuring complementary receptors, see Fig.~\ref{fig:Fig1}a. Both the colloidal probe particles ($2.12\pm0.06 \mu\text{m}$ silica spheres) and the glass surface are coated with a supported lipid bilayer (SLB) and functionalized with DNA linkers as the ligand-receptor system. Each DNA linker consists of a $77 $bp double stranded stem, which is modified with cholesterol on one $5\prime$ end to facilitate anchoring in the lipid membrane. Attached to the double stranded stem, the linkers feature a single-stranded overhang (sticky end) whose length and complementary base-pair sequence provide precise control over the tuneability of the hybridisation free energy. See Materials and Methods for details. The integration into a lipid membrane both on the colloidal probe and the surface provides full mobility of ligands and receptors if the lipid membrane is in the fluid state \cite{Rinaldin2019}.

To visualize and quantify the multivalent colloid-surface binding, we employ a combination of total internal reflection fluorescence microscopy (TIRFM) with Foerster Resonance Energy Transfer (FRET). We place fluorophores, which are also FRET pairs, on the $3\prime$ end of the complementary DNA linkers, see Fig.~\ref{fig:Fig1}a, and use dual color imaging with alternating laser excitation to investigate the DNA-DNA interactions in the colloid-surface contact area. The separate imaging of the channels provides information on the ligand and receptor distribution on the surface, see Fig.~\ref{fig:Fig1}b. Upon binding, the intensity of the ligand and receptor signal increases both on the colloid and surface, and, importantly, a FRET signal appears. See Fig. \ref{fig:Fig1}b. We verify that the FRET signal corresponds to the presence of a colloidal probe by overlaying it with a bright field image, see Fig.~\ref{fig:Fig1}b. The detection of the fluorescent and FRET signal is crucial to distinguish bound from unbound particles. The extended fluorescent patch in the contact area implies a local increase in the concentration of ligands and receptors, and originates from the recruitment of the surface-mobile DNA linkers to the contact area. The simultaneous appearance of the FRET signal indicates that there are multiple ligand-receptor interactions between the colloidal probe and the surface and shows that our system is multivalent. With this experimental setup, which combines tunable ligand and receptor densities, mobile binding sites and adjustable interaction strength (see Fig.~\ref{fig:Fig1}c), we can control and investigate the thermodynamic parameters relevant for superselective binding of multivalent colloids to a target surface and gain insights on the ligand-receptor dynamics on the molecular level.

\begin{figure*}[htb]
\centering
\includegraphics[width=.8\linewidth]{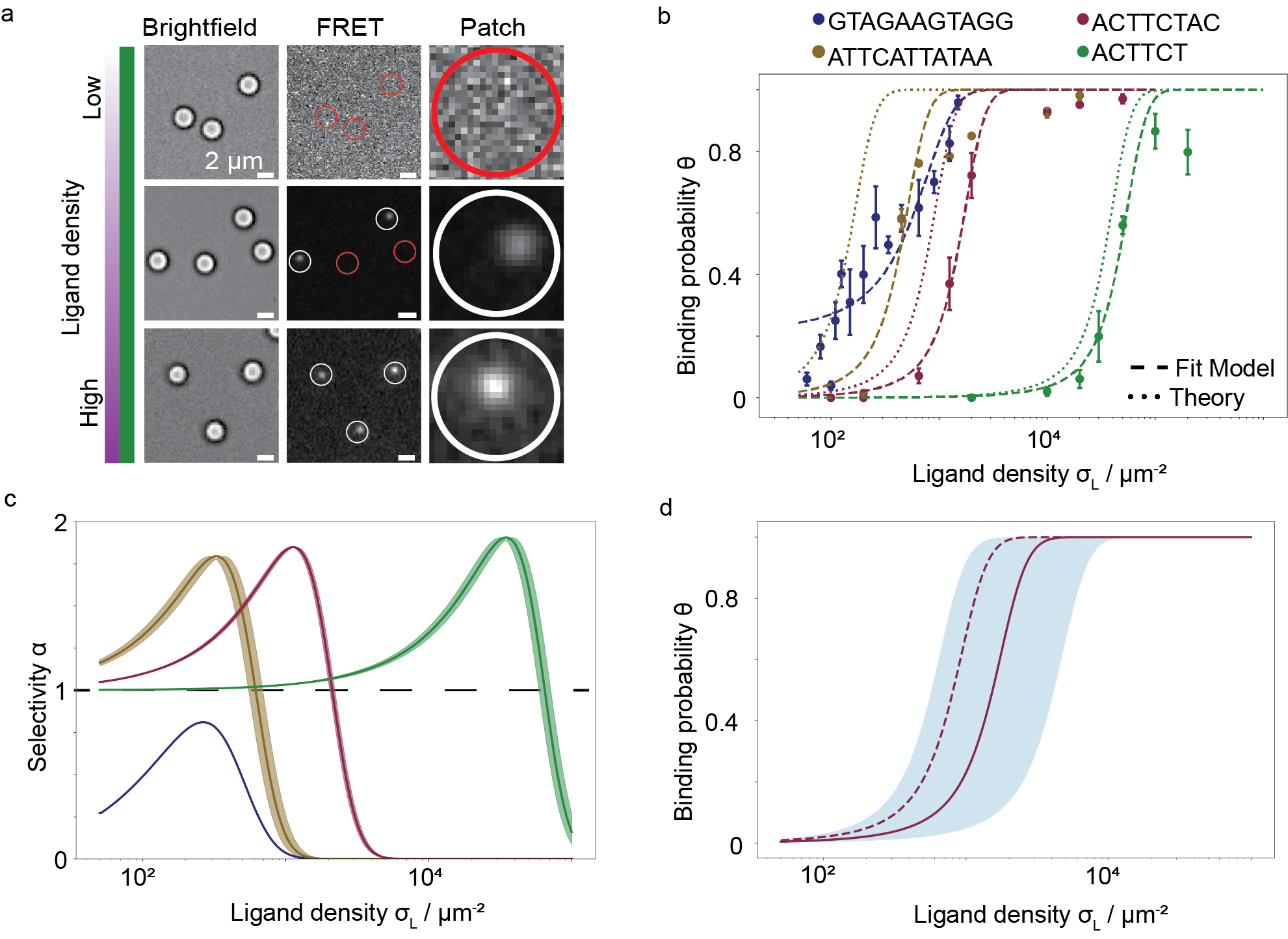}
\caption{\textbf{Superselective surface binding.} (a) Combining the information of the colloid position obtained from brightfield imaging with the FRET signal from bonded ligand-receptor allows us to discriminate whether a colloid-surface bond, a patch, has been formed. With increasing ligand density on the colloids larger and more bond patches are formed, implying an increase in the binding probability $\Theta$. The dynamic range between FRET images was individually adjusted to achieve an optimal visualization of the patch. 
(b) Measured colloid-surface binding probability $\Theta$ as a function of ligand density $\sigma_L$ for different ligand-receptor interaction strengths tuned by varying the length of the single-stranded end. Dashed lines are least-square fits to the model from Frenkel and co-workers \cite{Martinez-Veracoechea2011}; dotted lines represent the adsorption profile obtained using a theoretically computed $\Delta G_{\rm tot}$. The data with the longest sticky end GTAGAAGTAGC is fitted with a logistics function. Increasing binding strength shifts the curves to the lower ligand densities. The good agreement between fitting and theoretical evaluation is even more remarkable considering that errors in binding energies are greatly amplified. (c) The change in binding probability with ligand density can be measured by the selectivity parameter $\alpha = \frac{\text{d}\ln(\Theta)}{\text{d}\ln(\sigma_L)}$ and was derived from the fits in \ref{fig:Fig2} b). (d) Sensitivity of fitting model The solid and dashed line are the same as in b); the shaded region shows the error bar for an uncertainty of $\pm 1 k_B T$.}
\label{fig:Fig2}
\end{figure*}

The hallmark of superselective binding is a sharp, nonlinear change of the binding probability in a specific ligand or receptor density range. We start by investigating the colloid-surface binding probability with increasing ligand density, $\sigma_L$, while keeping the receptor density, $\sigma_R$,  on the surface fixed. Furthermore, we keep the overall DNA density constant on the colloidal probe by the addition of $77 $bp double stranded DNA that does not possess a sticky end. This ensures a constant concentration of cholesterol in the lipid membrane and hence constant membrane properties \cite{membrane,Zhang2020,Chakraborty2020}. At low $\sigma_L$ the fluorescent signal is homogeneously distributed over the colloid and the target surface and we do not observe fluorescent patches or a FRET signal (see Fig.~\ref{fig:Fig2}a). This indicates that binding does not occur despite the availability of DNA linkers on both colloid and surface. An increase of the ligand density on the colloid leads to the formation of patches on some probe particles, which implies that a fraction of the colloids in the sample are bound to the surface. Upon a further increase in $\sigma_L$  we observe that all colloids that are in close proximity to the membrane display a patch. The intensity of the patch differs between colloids, which is possibly due to the variability of ligand density between the functionalized colloids \cite{Delcanale2018}. The binding probability increases with the ligand concentration, however, superselectivity requires this transition to be non-linear.

Therefore, we determine the number of bound particles, $N_B$, relative to the total number of colloids, $N_C$, to measure the binding probability, $\Theta = \frac{N_B}{N_C}$. A value of $\Theta$ equal to $0$ implies that no colloids are bound, whereas the upper limit of $\Theta = 1$ is set by all colloids being bound to the surface. Besides varying the ligand density on the colloidal probe, we tested the binding behaviour for four different interaction strengths, i.e. for four different sticky ends. We find that the binding probability smoothly transitions from an unbound to a bound state and saturates at high ligand densities, shown in Fig.~\ref{fig:Fig2}b. The range of ligand densities where the transition occurs depends on the interaction strength: the higher the interactions strength, or, the longer the sticky end, the fewer ligands are required for binding the colloid to the surface. In addition, the slope of this transition becomes steeper for weaker ligand-receptor interactions, indicating a higher sensitivity of the binding probability to the ligand density. 

We examined the selectivity of the colloid binding for each binding probability curve by evaluating the relative change in binding probability with respect to a change in the ligand density $\sigma_L$ , also known as the selectivity parameter $\alpha = \frac{\text{d}\ln(\Theta)}{\text{d}\ln(\sigma_L)}$  \cite{Martinez-Veracoechea2011}. The system is superselective in a specific ligand density range if $\alpha \gg 1$. In order to evaluate $\alpha$ for the different sticky ends, we first need a mathematical description of $\Theta(\sigma_L)$. 

A physically-justified analytical form based on statistical mechanics considerations can be built by adapting a model first described by  Martinez-Veracoechea and Frenkel \cite{Martinez-Veracoechea2011}. In this model, the binding probability $\Theta$ is written as: 
\begin{align}
    \Theta = \frac{z q(N_L,N_R,G_{\text{bond}})}{1 + z q (N_L,N_R,G_{\text{bond}})},
\label{eq:theta}
\end{align}
where $z= \rho_B v_{\rm bind}$ is the multivalent particle activity in a (diluted) solution, $\rho_B$ being its bulk density and $v_{\rm bind}$ the binding volume, that is, the volume the particle centre of mass can move in while being able to form bonds to the surface. In this expression, a central role is that of $q(N_L,N_R,G_{\text{bond}})$, the ratio between the partition function in the bound and unbound state, which depends on the total number of ligands on the colloid and receptors on the surface, $N_L$ and $N_R$, respectively, as well as their binding (free)-energy, $G_{\rm bond}$. In our case, a simple mean-field approximation (see details in the supplementary information) leads to the formula:
\begin{align}
    q (N_L, N_R, G_{\rm bond})= \left[ 1 + N_R \exp\left(-\beta G_{\rm bond}\right) \right]^{N_L} - 1,
\label{eq:mean-field}
\end{align}
with $\beta = k_BT$ where $k_B$ is the Boltzmann constant and $T$ is the temperature.
In this model, a bound state is any state where at least one bond is present, as we measure in our experiments. The binding strength of the multivalent system is incorporated in $q$, which takes into account all possible binding configurations of the ligands and receptors, as well as information regarding the average strength of a single ligand-receptor bond, measured by $\exp\left(-\beta G_{\rm bond}\right)$.
Notably, in our system the bond strength is affected by experimental parameters such as the size of the rigid, double-stranded stem of the DNA, the sequence of the single-stranded part (sticky end) as well as DNA mobility on the colloid \cite{Varilly2012}. In particular, the latter introduces a dependence of the effective bond strength on the colloids area, as well as on that of the surface on which receptors are grafted \cite{Angioletti2014}. The exact value of $G_{\rm bond}$ can be calculated via detailed molecular simulations or experiments. Here, we leave it as a fitting parameter, and then compare the fitted value with an approximate theoretical expression derived by Mognetti et al \cite{mognetti2019} for mobile ligands and receptors and adapted here for our system. Within this framework, we obtain:
\begin{align}
    G_{\rm bond} = G_0 + G_{\rm conf} = G_0 + k_B T \log\left( 2 R_c A_{\rm tot} \rho^{\circ} \right),
\label{eq:bond}
\end{align}

where $G_0$ is the binding energy of the sticky end of the DNA in solution (which can be accurately estimated via Santalucia's nearest-neighbour rules \cite{Santalucia1998}) and $G_{\rm conf}$ is the so-called configurational contribution to the bond energy \cite{Varilly2012}. In our system, this last term turns out to be only dependent on the colloid radius, $R_c$, the total binding surface $A_{\rm tot}$, and the reference molar concentration $\rho^{\circ} = 1 M$, see details in the supplementary information. Importantly, using Eq. \ref{eq:bond} the binding probability $\Theta$ is fully determined, leaving no fitting parameter.

The results of fitting the experimental data on the binding probability as a function of the ligand density is reported in Fig.~\ref{fig:Fig2}, along with the experimental data, using both the expression where $G_{\rm bond}$ is left as a fitting parameter, which we will refer to as the free model, as well as the full theory. In the steep regime where $\alpha$ is determined, the free model nicely captures the experimental data for all sticky end sequences but the strongest one. For the strongest binding sequence ($\Delta G^0 = -17 k_BT$), however, the predicted trend is steeper than what is observed experimentally and for this reason the theoretical model cannot be relied on to calculate the value of the super-selectivity parameter $\alpha$. Thus we use an empirical logistic function with two parameters for this case, see Fig.~\ref{fig:Fig2}b.
The decrease in predictive power of the theoretical expression for increasing bond strengths is expected because the theoretical model is based on equilibrium considerations, where we assumed that binding and unbinding is fast enough for a colloid to fully sample all its possible binding configurations. However, this assumption becomes less and less justified as the bond lifetime increases, an increase expected to be exponential in terms of $G_{\rm bond}$. 

For the range of parameter in which the equilibrium model well describes the experimental data, we can compare the free model with the full theory, see Table \ref{tab:Tab1}. The agreement is remarkable, as the full theory provides a value for the only fitting parameter in the free model, the bond energy $G_{\rm bond}$, which only deviates from the one obtained through least-squares fitting by $\approx 1 k_BT$. Notably, even such a small discrepancy can result in a relatively large shift of the predicted adsorption probability because of the extreme sensitivity of the latter to $G_{\rm bond}$, as illustrated in Fig.~\ref{fig:Fig2}d, where the shaded region corresponds to the predicted curve obtained given an uncertainty of $1 k_BT$ on the bond energy.

Having fitted the data with an analytical form, we can easily compute the selectivity $\alpha$, see Fig.~\ref{fig:Fig2}c. Each sticky end shows a maximum selectivity for a specific ligand density. For the three sticky ends ATTCATTATAA ($\Delta G^0 = -13 k_BT$), ACTTTCTAC ($\Delta G^0 = -11 k_BT$) and ACTTCT ($\Delta G^0 = -7 k_BT$) we observe $\alpha \gg 1$, indicating that the colloid-surface binding is superselective. The longest and hence strongest-binding sticky end GTAGAAGTAGG ($\Delta G^0 = -17 k_BT$) however, does not exceed $\alpha = 1$ and is thus not superselective. The results of the binding probability for different ligand densities and interaction strength show the relevance of employing weak interaction to achieve superselective binding in multivalent systems, as previously pointed out by Martinez-Veracoechea et al. \cite{Martinez-Veracoechea2011}.

\begin{figure}[tbhp]
\centering
\includegraphics[width=0.8\linewidth]{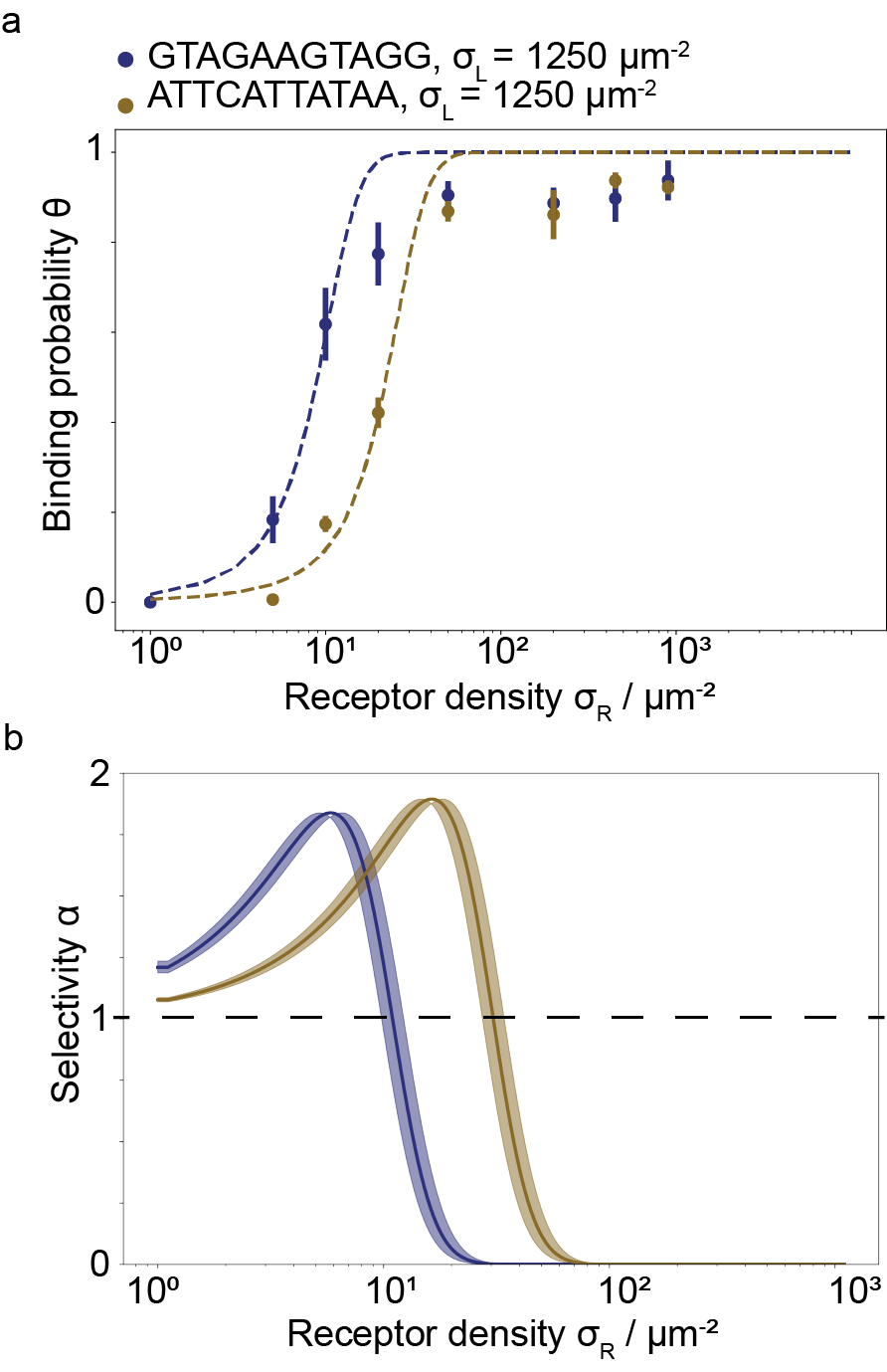}
\caption{\textbf{Multivalent binding as a function of the receptor density}. (a) Binding probability as a function of the receptor density for the $11$ bp sticky ends GTAGAAGTAGG and ATTCATTATAA at fixed ligand density of $1250 \mu\text{m}^{-2}$. (b) Selectivity parameter $\alpha$ resulting from a fit with the physical model shows a superselective regime with $\alpha > 1$ for both sticky ends.}
\label{fig:Fig3}
\end{figure}

\begin{table*}\centering
\caption{Comparison between the binding free energy obtained from the multivalent model and experimental data. The binding free energy $\Delta G_{\text{bond}}$ is obtained for the four sticky ends by computing the theoretical value ($\Delta G^{\text{theo}}$, see Methods) and by fitting Equation \ref{eq:theta} to the binding curves for a variation in the ligands (see Fig. \ref{fig:Fig2}), $\Delta G^{\text{fit}} \text{(lig)}$, and receptors (\ref{fig:Fig3}), $\Delta G^{\text{fit}} (\text{rec})$. The absolute differences between the experimental and theoretical values $\Delta G^{\text{fit}}$ - $\Delta G^{\text{theo}}$ are shown in the last column: the first value in each row is obtained using the experimental value of the free energy obtained varying the concentration of ligands $\Delta G^{\text{fit}} (\text{lig})$, while the second is computed using $\Delta G^{\text{fit}} (\text{rec})$; these differences are small with respect to thermal energy $k_BT$, verifying that our model is appropriate for our experimental setup.}

\begin{tabular}{cccccc}
Free energy ($k_BT$) & $\Delta G^{\text{0}}$  & $\Delta G^{\text{fit}} (\text{lig})$  & $\Delta G^{\text{fit}} \text{(rec)}$ & $\Delta G^{\text{theo}}$ & $|\Delta G^{\text{fit}}$ - $\Delta G^{\text{theo}}|$ \\
\midrule
ACTTCT & $-7$ & $32$ & ${\rm N/A}$ & $31$ & $1 - {\rm N/A}$ \\
ACTTCTAC & $-11$ & $28$ & ${\rm N/A}$ & $28$ & $0 - {\rm N/A}$ \\
ATTCATTATAA & $-13$ & $27$ & $25$ & $26$ & $1 - 1$\\
GTAGAAGTAGG &$-17$ & $27$ & $24$ & $21$ & $6 - 3$\\
\bottomrule
\end{tabular}
\label{tab:Tab1}
\end{table*}

Increasing the number of receptors on the flat surface changes the entropic effects upon binding and can increase the sensitivity of colloid-surface binding, similar to a change in ligand density. Here, we tune the selectivity of colloid-surface binding by changing the receptor density on the surface, while keeping the ligand density constant and at the same time investigating if the colloids indeed bind superselectively to a surface. We quantify the binding probability for two $11$ bp sticky ends ATTCATTATAA ($\Delta G^0 = -13 k_BT$) and GTAGAAGTAGG ($\Delta G^0 = -17 k_BT$) at a fixed ligand density of $1250 \mu\text{m}^{-2}$, see Fig.~\ref{fig:Fig3} a. Similar to a change in ligand density, we observe an increase of the binding probability until it saturates around $\Theta = 1$. 

We use the same physical model to evaluate the binding probability as well as the superselectivity parameter by fitting the experimental data, finding again good agreement between theory and experiments, see Fig.~\ref{fig:Fig3}a. We note that in this case the selectivity parameter for both $11$ bp sticky ends are larger than $1$ in a specific receptor density range, and thus both systems are behaving superselectively. Interestingly, in this case the sticky end GTAGAAGTAGG binds superselectively, whereas this was not observed in the system with fixed receptor density and varying ligand density. We suspect that a high valency on the colloid leads to a faster bond formation and thus equilibration of the system, which can explain this observation. These results indicate that the binding kinetics play an important role in multivalent bond formation.

For superselectivity to be observed on a given timescale, the dynamics of the individual bonds needs to be fast enough for them to behave reversibly on that timescale. In other words, bonds should constantly form and break. In fact, if bonds were irreversible the binding probability would be $1$ regardless of the density of ligands and receptors.

\begin{figure*}
\centering
\includegraphics[width=0.8\linewidth]{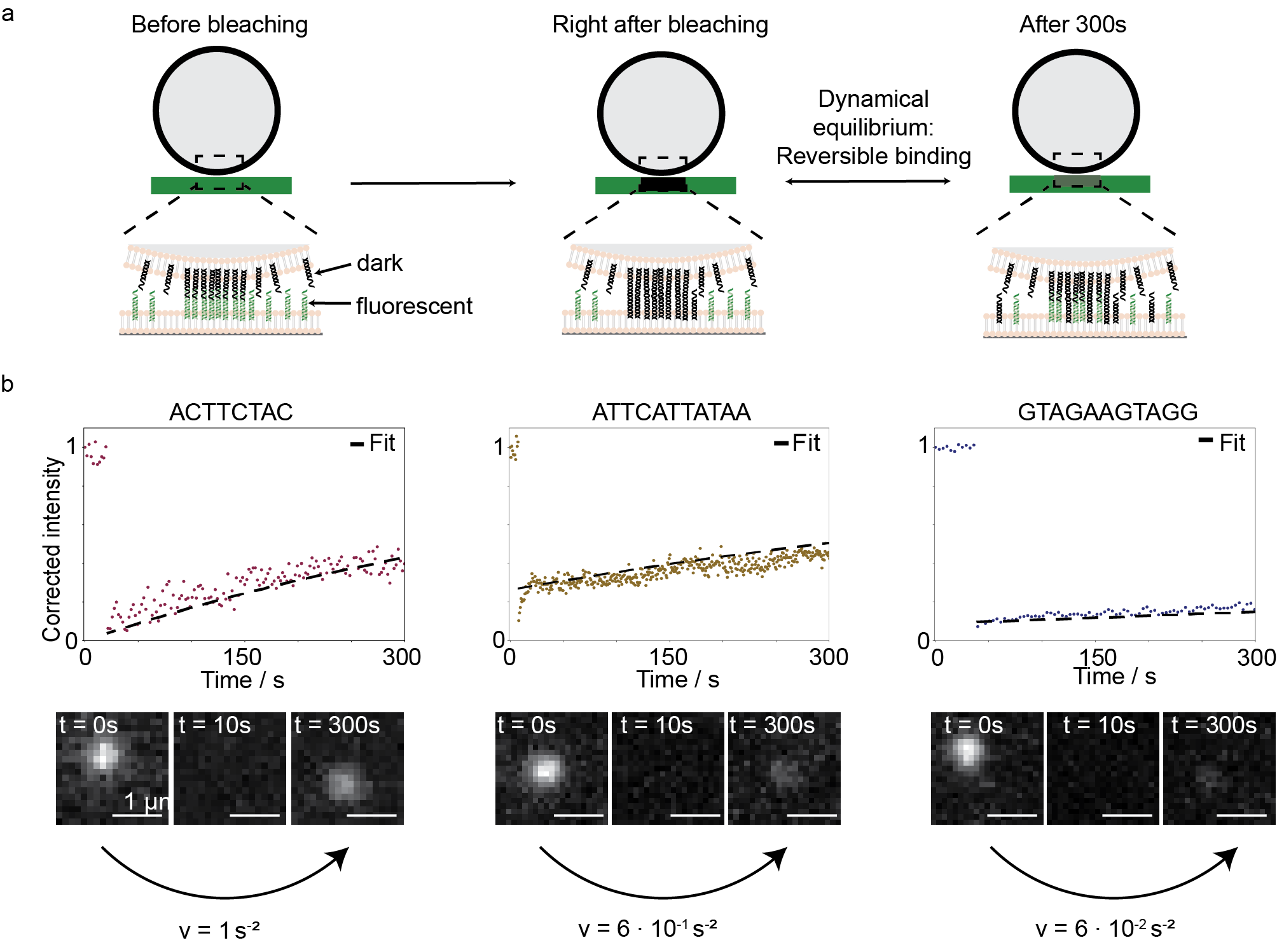}
\caption{\textbf{Ligand-receptor binding kinetics in a multivalent bond.} (a) Receptors and ligands with weak interactions repeatedly break and form bonds over time. We visualize their dynamic binding and unbinding by observing the fluorescent recovery after photobleaching (FRAP) of the receptors in the contact area of a colloidal particle and the target surface. For sufficiently weak ligand-receptor interactions the bonded pairs frequently break and bleached receptors can diffuse out of the patch with a flux $\propto k_{out}$. At the same time, unbleached receptors can diffuse into the patch with flux $\propto k_{in}$ and bind to ligands on the colloid. This exchange results in an increase of the receptor signal of the patch in time. (b) FRAP experiments for three sticky ends. Top row shows intensity recovery in time, bottom row shows the recovery of the fluorescent intensity in the receptor channel in time.  Colloidal probes coated with the sticky end ACTTCTAC have a ligand density of $100.000 \mu \text{m}^{-2}$ and colloids coated with the 11bp sticky end comprise of $10.000 \mu \text{m}^{-2}$ ligands. The target surface was functionalized with a receptor density of $450 \mu \text{m}^{-2}$.  Fitting the FRAP curves with Eq.~\ref{eq:intensity}, \ref{eq:kin}, \ref{eq:kout} we can quantify the initial speed of recovery, which is defined as the product $\nu = \sigma_0 k_{\text{out}} k_{\text{in}}$.}
\label{fig:Fig4}
\end{figure*}

Since our setup allows for the direct visualization of the spatial receptor distribution on the flat surface, we can visualize the exchange of the receptors inside the contact area with unbound receptors in close proximity of the patch using Fluorescent Recovery After Photobleaching (FRAP), see Fig.~\ref{fig:Fig4}a. We performed  bleaching experiments of ligand-receptor patches in the contact area for three sticky ends ACTCTAC ($\Delta G^0 = -11 k_BT$), ATTCATTATAA ($\Delta G^0 = -13 k_BT$) and GTAGAAGTAGG ($\Delta G^0 = -17 k_BT$) and recorded the signal recovery up to $300 $s, see Fig.~\ref{fig:Fig4}b. After bleaching the receptors in the patch, we observe a recovery of the signal for all sticky ends, albeit at different rates. Furthermore, the recovery rate depends on the sticky end: the stronger the hybridization free energy, the longer the recovery of the receptor signal takes. This shows that the hybridization energy of the individual receptors and ligands influences the timescale on which the formation and dissolving of bonds occur.

\begin{figure}
    \centering
    \includegraphics[width=9cm]{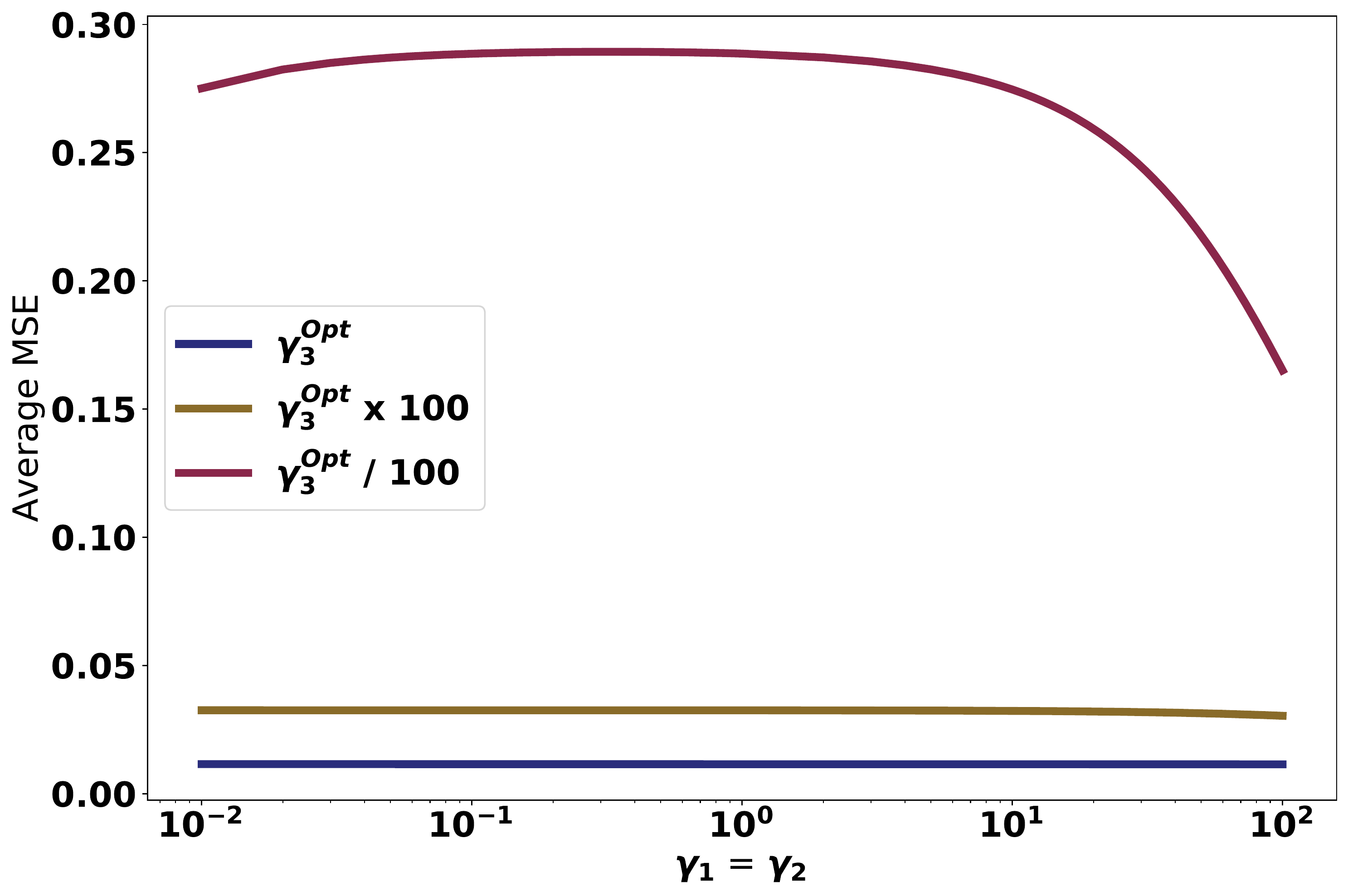}
    \caption{Mean square error of the fit as a function of  $\gamma_1 (=\gamma_2)$, shown here for both the optimal value of $\gamma_3$ (blue line) as well as for two sub-optimal values (yellow and purple line). As it can be seen, the mse of the fit is almost insensitive to the value of $\gamma_1$ chosen compared to the effect of $\gamma_3$. Because $\gamma_1 = \gamma_2$ are related to the diffusion time of ligands whereas $\gamma_3$ to the bond breaking time, the observed insensitivity is what one can expect when kinetics is dominated by the latter process.}
    \label{fig:Fig9}
\end{figure}

Mathematical modelling allows us to unravel some details of the kinetics in our system and gauge the relative importance of different processes in the observed behaviour.
At a coarse-grained level, the intensity recovery over time for our system should be well-described within the framework of Langmuir kinetics \cite{aziz}, whose assumptions we use here to derive a simple two-component model (corresponding to the description of bleached vs unbleached DNA) to gain a better, quantitative understanding of the dynamics of receptor-ligand bonds in the patch. Within this model the normalised intensity $I$ as a function of time after bleaching is given by (see the Supplementary Information for details):
\begin{align}
    I = \frac{ k_{\rm in} \sigma_0 + k_{\rm out} e^{-k_{\rm in} \sigma_0 t} - k_{\rm in} \sigma_0 e^{-k_{\rm out} t}} {k_{\rm out} + k_{\rm in} \sigma_0},
\label{eq:intensity}
\end{align}
where $k_{\text{in}}$ and $k_{\text{out}}$ are two kinetic constants related to the inward and outward flux of receptors from the bulk to the patch and vice-versa, respectively, and $\sigma_0$ is the bulk density of receptors on the surface. 
The initial speed of recovery can be approximated as proportional to the product $k_{\text{in}} k_{\text{out}}$ (again, see the supplementary information for details). The exact relation between microscopic details of the system and the value of $k_{\text{in}}$ and $k_{\text{out}}$ is difficult to quantify without the use of accurate molecular simulations. For example, the speed at which receptors can diffuse will be dependent on the viscosity of the lipid membrane and the interaction between cholesterol linkers and lipid chains. Here, instead, we therefore only provide an approximate expression showing how the kinetic constant is expected to depend on parameters such as the receptors diffusion coefficient in the lipid membrane $D$, and the sticky end hybridization free energy energy $G_0$, which reads:
\begin{align}
k_{\text{in}} &= \gamma_1 D \label{eq:kin}\\
k_{\text{out}}^{-1} & = k_{\text{diff}}^{-1} + k_{\text{break}}^{-1} =
\gamma_2 R_c L / D + \gamma_3 \exp(-\beta G_0 )\label{eq:kout}
\end{align}

where $\gamma_i, i=1,2,3$ are system-dependent constants within our model, which we will use to fit our experimental data and $L$ is the length of the receptor. We obtained these approximate formulas by assuming that $k_{\rm in}$ depends on the rate at which ligands from the outside of the binding patch diffuse into it, whereas $k_{\rm out}$ arises from a two-step process, where the bound ligand-receptor pairs in the patch first unbinds, and later the receptor diffuses out of the binding patch, whose size will be of the order of $R_c \: L$ (see Supplementary Information for details). Given their origin from the solution of a diffusion problem, and based on dimensional analysis, $\gamma_1$ and $\gamma_2$ should be non-dimensional coefficients of order one and only dependent on geometrical factors, i.e., the specific boundary conditions under which the ligands evolve. Because the sticky ends are located at the top of a long dsDNA stem, they will be mostly far from the lipid membrane and we do not expect them to strongly influence the diffusion coefficient of different ligands and receptors. For this reason, we fix $\gamma_1 =\gamma_2=1$ for all the systems investigated. In contrast, $\gamma_3$ has the dimensions of a frequency and can be thought of as the natural frequency at which a bond of $G_0 = 0$ will break. Because $\gamma_3$ is expected to be a function of the exact microscopic details of the bond-breaking mechanism, we leave it as a system-specific fitting parameter. Thus $k_{\rm in}$ will be constant by construction for all ligands and receptors, while we expect $k_{\rm out}$ to decrease as the bond between the sticky ends of the DNA becomes stronger, due to its dependence on $G_0$.\\
A fit of Eq.~\ref{eq:intensity} to the intensity recoveries yields the unknown parameter $k_{\text{out}}$, see Fig.~\ref{fig:Fig4}b. As expected, the product $k_{\text{out}}k_{\text{in}}$ which quantifies the initial speed of recovery shows a decrease with increasing bond strength. A more detailed analysis of the fitting results provides two crucial physical insights: the first is, that the quality of the fitting is essentially insensitive to the value of $k_{\rm in}$, whereas it strongly depends on $k_{\rm out}$, see Fig.~\ref{fig:Fig9}. This suggests that a correct description of the process by which receptors diffuse out of the binding patch is more important then diffusion towards it. To make this more quantitative, we calculate the non-dimensional parameter $\delta = \frac{k_{\rm in}\sigma_0}{k_{\rm out}}$, whose values are reported in Tab.~\ref{tab:kin-kout}, as a measure of the relative magnitude of the ingoing vs outgoing flux towards the patch. For all sequences used here, $\delta \gg 1$, showing that the outward diffusion is much slower than inward diffusion, implying that it is the bottleneck of the signal recovery process. 

Secondly, we find that  $k_{\text{diff}} \gg k_{\text{break}}$. Interpreted in light of the model expressed in Eqs.~\ref{eq:kin} and\ref{eq:kout}, this suggests that the kinetics in our system is dominated by bond breaking rather than diffusion. In other words, bond breaking is the rate-limiting factor for the system to sample different binding configurations. This an important finding since super-selectivity has been shown to arise from statistical mechanical effects, in particular, the steep increase in the combinatorial entropy of binding as the number of ligands and receptors increases. To observe an ergodic sampling of these configurations within the timescales accessible to experiments, and thus observing superselective behaviour (which is an equilibrium property), requires the system dynamics to happen on much shorter timescales. Here we have shown that, at least for our system, these are dominated by the unbinding kinetics of the ligand-receptor pair.

\section*{Conclusion}

Multivalent binding in a fully mobile system is a highly dynamic process that can show superselective surface binding at the right ratio of enthalpic and entropic contributions. In this work, we have combined experiments with theory to investigate the binding probability of multivalent ligand and receptor interactions between two surfaces that provide full mobility to the ligands and receptors. Our experimental setup allowed us to directly visualize the spatial distribution of the individual binding sites. We have shown that multivalent binding between fluid surfaces is characterized by receptor and ligand clustering as suggested by Lanfranco et al. \cite{Lanfranco2019} and Dubacheva et al. \cite{Dubacheva2019}. Following the design rules proposed by Frenkel et al. \cite{Martinez-Veracoechea2011} we demonstrated that we can achieve superselective binding by tuning  the hybridisation energy of the individual ligand and receptors and measured the effective free energy of binding. Finally, we visualized and quantified the reversibility of weak bonds through the highly dynamic exchange of bound receptors with unbound ones from outside the binding patch.

Future studies with this model system can provide exciting insights into the binding kinetics of multivalent interactions at the ligand-receptor level, for example the formation of bonds and development of the spatial distribution in time. The observed timescale of bond formation and importance of the time scale for bond breakage might be crucial for competitive binding of various receptors on cell surfaces. Moreover, our experimental system can provide insights into membrane deformations caused by locally high receptor and ligand accumulations, which is important for the initiation of endocytosis. Surface targeting in biological systems is often governed by more then one type of receptors and ligands, and our model system can straightforwardly be extended to study the effect of competing interactions between more types of ligands and receptors on superselectivity\cite{Curk2017,Bray1998,Stefanick2019}. Finally, our experimental system might provide key insights for applications in nanomedicine as it can be used to improve specific target binding while reducing off-target binding, useful for example for drug-delivery.

\section*{Materials and Methods}

\paragraph*{{\normalfont \textbf{DNA strands.}}}
All DNA strands (Integrated DNA Technologies, Inc., Eurogentec, IBA) with a sticky end are functionalized with cholesterol at the $5\prime$. The $3\prime$ is modified with a fluorophore ($Cy3$/$Cy5$, $Cy3/Atto655$). The complementary backbone DNA strand had a length of $77$ bp and cholesteryl-TEG at the $3\prime$ end. Single stranded DNA with the sticky end and single stranded backbone were annealed to $95^\circ$C and slowly cooled in $0.2^\circ$C/minute steps in TAE-NaCl (TAE, $100 $mM NaCl, $\text{pH} = 8$) buffer in a Thermocycler. The resulting DNA strands are double-stranded with a double cholesterol anchor and a single stranded overhang. The hybridized DNA strands were stored in TAE-NaCl buffer at $4^\circ$C. The single stranded overhang varies in sequence and length with a hybridisation energy ranging between $\Delta G^0 = -7 k_B T -(-17 k_B T)$ (DinaMeltServer Reference). The sticky ends used as the ligand and receptor system are ACTTCT, ACTTCTAC, ATTCATTATAA, GTAGAAGTAGG and their respective complementary sequences. 
\paragraph*{{\normalfont \textbf{DNA coated Colloid supported lipid bilayers.}}}
We coated commercial silica spheres (Microparticle GmbH) of $2.12 \mu$m with a supported lipid bilayer (SLB). To do so, we mixed the silica particles ($0.5 \text{wt}\%$) with small unilamellar vesicles (SUVs) consisting of the desired lipid composition and incubated the mixture at room temperature for $30$ min. To obtain the SUVs, we first added the desired volume of $18:1$ DOPC lipids (Avanti Polar Lipids, stored in chloroform) into a glass vial and let it dry overnight in a vacuum desiccator. Next, we resuspended the dried lipids in TAE-NaCl buffer and  extruded the solution with an Avanti mini extruder through a membrane with pore size of $30$ nm yielding a transparent solution. By mixing the SUVs with the colloids, the SUVs spread on the colloid surface to form a supported lipid bilayer (SLB). Excess SUVs were removed by centrifugation of the mixture at $2000$ rcf for $30$ s and subsequent replacement of the supernatant with fresh TAE-NaCl buffer. The desired concentration of hybridized DNA was added to the colloid supported lipid bilayers (CSLB) and incubated for 1h at room temperature \cite{VanDerMeulen2013,Rinaldin2019}. Using the stock concentration and surface area of the colloids in solution, we estimated the final DNA surface density $\sigma_{\text{DNA}}$ on the CSLB, which typically ranged between $60\mu\text{m}^(-2)-100.000 \mu\text{m}^(-2)$. After the incubation time we washed the mixture trice by centrifugation at $2000$ rcf for $30$ s and replacement of the supernatant with fresh buffer. The last replacement of the supernatant was done with imaging buffer ($0.8\%$ dextrose [Sigma], $1$ mg/mL glucose oxidase [Sigma Aldrich], $170 $mg/mL catalase [Merck], and $1$ mM Trolox [$(\pm)-6-\text{hydroxy}-2,5,7,8-\text{tetramethylchromane}-2- \text{carboxylic acid}$, $238813$] [Sigma Aldrich]\cite{VanGinkel2018}) to reduce the bleaching of the fluorophores during imaging.
\paragraph*{{\normalfont \textbf{DNA functionalized supported lipid bilayer on flat glas surface.}}}
The microscopy slides and coverslips were sonicated for $30 $min each in $2\%$ Hellmanex, acetone ($>99.9\%$) and potassium hydroxide solution (KOH. $1 $M). Between each change of chemical, the glassware was rinsed with milliQ water and blown dry with nitrogen before use. The experiments were performed in a flow channel consisting of parafilm slices between a glass microscope slide [VWR] and a glass coverslip [VWR]. Placing the construct on a heating stage at $125^\circ$C melted the parafilm and bound the objective slide and coverslip together yielding four $2\text{mm} x 24\text{mm}$ flow channels. Before we injected SUVs into the flow channels, we cleaned the channels with TAE-NaCl buffer. After $30$ min we washed out the excess SUVs with TAE-NaCl buffer and added  DNA with the complementary DNA sequence at the desired concentration with respect to the DNA CSLB. Here, the resulting DNA surface densities $\sigma_\text{DNA}$ used varied between $5\mu\text{m}^(-2)-1500 \mu\text{m}^(-2)$. After $1$ hour incubation, we flushed the channels three times with TAE-NaCl buffer before injecting the DNA CSLBs to the flow channels.
\paragraph*{{\normalfont \textbf{Total Internal Reflection Microscopy.}}}
The colloidal silica particles quickly sedimented to the flat SLB due to their density being higher than that of water. To image the colloid-membrane interactions we used Total Internal Reflection Fluorescence Microscopy (TIRF) on an inverted fluorescence microscope (Nikon Ti2-E) upgraded with an azimuthal TIRF/FRAP illumination module (Gataca systems iLAS 2) equipped with a $100x$ oil immersion objective (Nikon Apo TIRF, 1.49NA). To investigate the DNA-DNA interactions in the colloid-surface contact area, we used dual color imaging with alternating laser excitation with wavelength $561$ nm and $640$ nm (Cairn Research Optosplit II ByPass, EM-CCD Andor iXON Ultra 897). This technique allowed for alternating excitation of the donor and receptor, yielding a Foerster Resonance Energy Transfer (FRET) when the two complementary DNA linker strands hybridized. Subsequently, we acquired a Brightfield image to localize the colloids on the surface (CCD Retiga R1). This setup was also used to perform Fluorescence Recovery after Photobleaching (FRAP) experiments to investigate the mobility of DNA in a membrane and patch. Per sample we imaged 100 colloids in at least two independent experiments. The errorbar on the binding probability represents the standard error of the mean. The error bar on the selectivity $\alpha$ results from the least square fitting error.
\paragraph*{{\normalfont \textbf{Image Analysis.}}}
After the acquisition of the data we deinterleaved and cropped the resulting images with respect to the three fluorescent channels corresponding to the donor, acceptor and FRET. For each measurement, we first acquired an image with fluorescent beads, which is used for the spatial calibration of the channels and the Brightfield image. An ImageJ plugin \cite{Preibisch2010} was used to overlay the fluorescent channels with the Brightfield image to locate the DNA-DNA interactions with respect to the colloids. We defined a colloid bound to the flat surface via DNA-DNA interactions if we could visually differentiate the signal in the colloid-surface contact area of the donor, acceptor and FRET channel from the local background signal.\\
For the FRAP experiments we extracted the intensity profile of the bleached area via ImageJ and first subtracted the background noise of the microscope and then normalized the raw intensity with respect to the initial unbleached intensity. Per condition we imaged at least 3 patches and in 3 independent experiments. 
\paragraph*{{\normalfont \textbf{Statistical mechanics model of adsorption Multivalency.}}}
We implement a model for the adsorption probability $\Theta$ of colloids to a receptor decorated surface through multivalent interactions by combining various results previously derived in the literature. In particular, we take $\Theta$ to have the form in Eq.\ref{eq:theta}, as shown e.g. in \cite{Martinez-Veracoechea2011,Dubacheva2015}, where $z = \rho_B v_{\rm bind}$, $\rho_B$ being the bulk density of colloids and $v_{\rm bind}$ their binding volume. For what concerns the value of the partition function $q$, we make a mean-field approximation, that is, we assume that the binding of a ligand to a certain receptor does not strongly affect the possibility for another ligand to bind it, while at the same time maintaining the constraint that a ligand cannot bind more than one receptor at any one time (the so-called valence-limited condition \cite{Varilly2012}). As previously discussed \cite{Varilly2012,Angioletti2013}, this mean-field approximation works well for weak ligand-receptor bonds and when the number of receptors is much larger than the number of ligands. In our system, both conditions are satisfied as the effective bond energies are very large and positive while the number of receptors is around 5 orders of magnitude larger than the maximum number of ligands used.\\

Under these assumptions we can write the ratio between the bound and unbound partition function for a whole colloid as \cite{Martinez-Veracoechea2011}:

\begin{equation}
    q = (1 + N_{\rm R} e^{- \beta G_{\rm bond}})^{N_{\rm L}} - 1
    \label{eq:q}
\end{equation}

 where the minus one account for the fact that colloid is considered bound if at least 1 bond is present. In Eq.~\ref{eq:q} $G_{\rm bond}$ is the effective bond energy, including a configurational entropic contribution \cite{Varilly2012}. Care must be taken in the interpretation of $N_{\rm R}$ and $N_{\rm L}$ in this formula. For colloids with surface-grafted ligands and SLBs with grafted receptors, these would be the number of ligands and receptors, respectively, that are at available for binding given a certain orientation of the colloid \cite{Martinez-Veracoechea2011}. However, because ligands and receptors are mobile in our system, $N_{\rm R}$ and $N_{\rm L}$ correspond to the total number of receptors on the adsorption surface and the total number of ligands per colloid. Thus, in our case, $N_{\rm L}$($N_R$) can be approximated as the average ligand (receptor) grafting density $\sigma_L(R)$ multiplied by the colloid (binding surface) area.\\

In order to obtain a theoretical estimate of the free-energy of a single ligand-receptor bond $G_{\rm bond}$, we use the formula \cite{Varilly2012}:

\begin{equation}
    G_{\rm bond} = G_0 + G_{\rm conf}
    \label{eq:bondg}
\end{equation}

where $G_0$ is the nucleotide-sequence-dependent free-energy of binding in solution for the free sticky end of the complementary DNA, which we calculated using Santalucia's nearest neighbour model \cite{Santalucia1998}. As explained in detail in \cite{Varilly2012}, $G_{\rm conf}$ is a configurational entropic penalty due to the restriction of the phase space for the ligand and receptor pair upon binding. This configurational contribution is evaluated for mobile, rod-like ligands, freely pivoting around their tethering point and with point-like sticky-ends, exactly the conditions of our system where ligands and receptors comprise a long and rigid dsDNA part to which a short sticky end of ssDNA is attached. In particular, $G_{\rm conf} = G_1 + G_2 + G_3$ in our system, where the three different contributions are:

\begin{enumerate}
    \item $G_1 = -k_B T \log\big(\frac{A_p}{A_{\rm tot}}\big)$. This is the entropic cost of localising a 
    receptor within a patch of area $A_p = 2\pi R_c L$ (see e.g., \cite{Tian2020}) where binding with a ligand can occur. Here, $R_c$ is the radius of the colloid and $L$ the length of the receptor, because such receptor would otherwise be able to span the whole adsorption surface ($A_{\rm tot}$) when in the unbound state. 
    \item $G_2 = -k_B T \log\big(\frac{A_p}{A_c}\big)$. Similarly to $G_1$, this contribution represent the entropic cost of localising a ligand from the whole colloid surface ($A_{c}$) to the patch area.
    \item $G_3 = k_B T \log(\rho_0 L A_p)$. This latter term is the entropic cost of bonding a rod-like receptor and a rod-like ligand together, assuming a fixed particle-surface distance equal to the ligand/receptor length $L$ (see Ref.~\cite{mognetti2019} for details). 
\end{enumerate}

Combining the three contribution and with some elementary algebra we obtain:

\begin{equation}
    G_{\rm conf} = + k_B T \log(2 R_c A_{\rm tot} \rho_0)
    \label{eq:bond-final}
\end{equation}

Interestingly the final expression for the configurational free-energy does not depend on the length of the ligands/receptor or the contact area, implying that it will have the same value for all of the systems considered in our work. Finally, we note that the mean-field calculation of the partition function and that of the configurational bond energy $G_{\rm conf}$, taken together, turn out to be exactly equivalent to a formula for the adsorption free-energy $F_{\rm ads}$ by Mognetti {\it et al} in Ref.~\cite{mognetti2019}. To connect the seemingly different treatments, one only needs to recognise that:

\begin{equation}
    \exp(-\beta F_{\rm ads}) - 1 = q,
    \label{eq:connection}
\end{equation}

where the $-1$ ( as discussed in \cite{Angioletti2017}) stems from the fact that we consider as bound only colloids where at least a single bond is present, whereas $F_{\rm ads}$ is calculated using the zero-energy reference as a state with no bonds.\\

Substituting our experimental parameters in Eq.~\ref{eq:bond-final}, we get a fixed value of $G_{\rm conf} \approx 38.7 k_B T$ that once summed with the hybridisation free energies $G_0$ computed with DINAMelt allows us to predict the fitted values of $G_{\rm bond}$ in our system with an average accuracy of less then $\pm 1.5 k_B T$.

\paragraph*{{\normalfont \textbf{Kinetics of the Photobleaching experiments.}}}
\begin{table*}\centering
\caption{Parameters of the kinetic model for multivalent recovery after photobleaching described in Eq.~\ref{eq:intensity}, \ref{eq:kin}, \ref{eq:kout} fitted to experimental data. As detailed in the main text, $\gamma_3$ was the only optimised fitting parameter; $\gamma_1$ and $\gamma_2$ values were fixed to 1, while the diffusion coefficients employed for the computation of the reported quantities were obtained experimentally. In addition to the optimal value of $\gamma_3$, various quantities describing the relative importance of different events in the overall kinetics of the system are showed: $k_{\text{in}} \sigma_0$ and $k_{\text{diff}}$ specify the relevance of diffusion in determining the inward and outward flux; $k_{\text{break}}$ describes the speed of the bond-breaking step involved in diffusion exiting the patch; $k_{\text{out}}$ is the overall kinetic constant for the outgoing flux; the product $\nu = \sigma_0 k_{\text{out}} k_{\text{in}}$ quantifies the initial speed of recovery; finally $\delta = \frac{k_{\text{in}} \sigma_0}{k_{\text{out}}}$ is a non-dimensional parameter providing an estimate of the relative magnitude of the ingoing vs outgoing flux. From the values shown it can be immediately seen how the limiting step of our system dynamics is represented by the bond breaking event ($\delta \gg 1$ for all DNA sequences) which becomes progressively slower as the hybridization free energy increases. The irrelevance of the speed of diffusion has been further confirmed by the results of the parameter space analysis we conducted (see Fig.~\ref{fig:Fig9}) which showed how the quality of the fitting is negligibly affected by the value of the parameters $\gamma_1$ and $\gamma_2$ (included in the expressions for $k_{\text{in}}$ and $k_{\text{diff}}$). 
}
\begin{tabular}{cccccccc}
DNA sequence &  $\gamma_3 \left[s^{-1} \right]$ & $k_{\text{out}} \left[s^{-1} \right]$  & $k_{\text{in}} \sigma_0 \left[s^{-1} \right]$  & $k_{\text{diff}} \left[s^{-1} \right]$ & $k_{\text{break}} \left[s^{-1} \right]$  & $\nu \left[s^{-2} \right]$ & $\delta$  \\
\midrule
GTAGAAGTAGG & $1.452 \cdot 10^{-4}$ & $2.293 \cdot 10^{-4}$ & $270$ & $17.241$ & $2.293 \cdot 10^{-4}$  & $6.192 \cdot 10^{-2}$ & $1.178 \cdot 10^{6}$ \\
ATTCATTATAA & $2.378 \cdot 10^{-3}$ & $1.335 \cdot 10^{-3}$ & $450$ & $28.736$ & $1.335 \cdot 10^{-3}$ & $6.008 \cdot 10^{-1}$ & $3.370 \cdot 10^{5}$ \\
ACTTCTAC & $9.178 \cdot 10^{-3}$ & $1.871 \cdot 10^{-3}$ & $720$ & $45.977$ & $1.871 \cdot 10^{-3}$ & $1.347 \cdot 10^{0}$ & $3.847 \cdot 10^{5}$ \\
\bottomrule
\end{tabular}
\label{tab:kin-kout}
\end{table*}

We provide here details of the derivation of a simple mathematical model to describe the photobleaching / recovery experiments. We start by assuming a Langmuir-type dynamics to describe the population of bleached DNA ($N_2$) and the ``unbleached one'' ($N_1$) inside the binding patch, i.e. the region where ligand-receptor bonds between a colloid and the surface are present. In practice, this means considering that the patch comprises $N$ sites, that initially all are occupied by bleached receptors. Because of excluded volume interactions, no two receptors can reside on the same site. Given the previous description we have the following boundary conditions in time:

\begin{align}
N_2( t = 0 ) &= N_{\rm max}\\
N_1( t = 0 ) &= 0
\label{eq:boundary}
\end{align}

where $N_{\rm max}$ is the maximum amount of DNA that can be inside a patch in equilibrium. Furthermore, as in Langmuir kinetics we assume that the fluxes $dN_i/dt, i = 1,2$ of bleached and unbleached DNA are both linear in the amount of free sites and in the amount of DNA outside of the patch that could replace them (for the incoming flux inside the patch) and linear with respect to the amount of DNA already in the patch (for the outgoing flux). The terms ``ingoing'' and ``outgoing'' simply refer to the flux from the outside to the patch region, and vice versa. We thus have the following set of linear coupled differential equations:

\begin{align}
\frac{dN_1}{dt} &= + k_{\rm in} ( N_{\rm max} - N_1 - N_2 ) \sigma_{\rm 1,out} - k_{\rm out} N_1\\
\frac{dN_2}{dt} &= + k_{\rm in} ( N_{\rm max} - N_1 - N_2 ) \sigma_{\rm 2,out} - k_{\rm out} N_2,
\end{align}

where $k_{\rm in}$ and $k_{\rm out}$, both positive, are the two proportionality constant for the flux and $\sigma_{1,\rm out}$, and $\sigma_{2,\rm out}$ are the density of unbleached and bleached DNA outside of the patch. In principle, these quantities are also dynamical quantities changing in time. However, because the total amount of DNA outside the patch is much larger than that inside, we can just assume them to be constant and set $\sigma_{1,\rm out}(t=0) \approx \sigma_0$ and, for the same reason, $\sigma_{2,\rm out}(t=0) = 0 $. Hence we can simplify the previous equations as:

\begin{align}
\frac{dN_1}{dt} &= + k_{\rm in} ( N_{\rm max} - N_1 - N_2 ) \sigma_0 - k_{\rm out} N_1\\
\frac{dN_2}{dt} &= - k_{\rm out} N_1
\end{align}

which with the boundary conditions Eq.\ref{eq:boundary} have the simple solution:

\begin{align}
f_1 &= \frac{N_1(t)}{N_{\rm max}} = \frac{ k_{\rm in} \sigma_0 + k_{\rm out} e^{-k_{\rm in} \sigma_0 t} - k_{\rm in} \sigma_0 e^{-k_{\rm out} t}} {k_{\rm out} + k_{\rm in} \sigma_0} \\
f_2 &= \frac{N_2(t)}{N_{\rm max}} = e^{-k_{\rm out} t} 
\end{align}

Note that if one normalises the signal from the patch with respect to the value it would have if all sites where occupied by unbleached DNA, this normalised signal is exactly given by $f_1$. 

Signal recovery by replacing bleached DNA with unbleached one requires bonds to be reversible and in dynamical equilibrium. Here, we quantify how fast this occur by looking at the initial speed of recovery, that is, the initial flux of unbleached particles towards the patch, which is given by a Taylor expansion of $dN_1/dt$, which to leading order is:

\begin{equation}
v_{\rm rec} = \frac{dN_1}{dt} \approx \frac{1}{2} k_{\rm in} k_{\rm out} \sigma_0 t + \mathcal{O}( t^2 )
\end{equation}

Because of their dependence on molecular details of the system, determining the exact value of $k_{\rm in}$ and $k_{\rm out}$ requires in principle expensive molecular simulations, outside the scope of this work. In the theory we present here we simply provide and fit approximate expressions for $k_{\rm in}$ and $k_{\rm out}$ to characterise the dynamics of the system; this allows us to provide an estimate of their scaling as a function of some fundamental microscopic parameters of the system. Let us first deal with $k_{\rm out}$. In this Langmuir model, this quantity is proportional to the frequency with which a DNA inside the patch escapes from it. Let us call $D$ the diffusion coefficient of DNA on the surface and assume that the escape time $t_{\rm out} = 1/k_{\rm out}$ can be written as the sum of two processes:

\begin{enumerate}
    \item Unbinding of the DNA from a potential partner, which takes an average time $\tau_{\rm unbind}$. If we consider this as a bond-breaking reaction, we expect this time to be exponentially dependent on the hybridisation energy in solution of the DNA sticky end, $G_0$ in the main text, hence $\tau_{\rm unbind} = \gamma_3 \exp(-\beta G_0)$, $\gamma_3$ having the dimension of a frequency. In practice, one can expect $\gamma_3$ to be dependent on the microscopic details of the intermediate steps along the unbinding pathway, and thus have a certain dependence on the sequence of nucleotides use \cite{unbinding}.
    \item Diffusion of the unbound DNA outside the patch area, which takes a time $\tau_{\rm diff} \approx \gamma_2'  A_p^* / D = \gamma_2  R_c L / D$, where this time $\gamma_2$ is a non-dimensional constant, approximately of order 1, where we collect the effects of all the finer details of the diffusion process. $R_c$ is the colloid radius and $L$ the length of the receptor.
\end{enumerate}

From these considerations, we obtain the final formula:
\begin{equation}
    k_{\rm out}^{-1} =  \gamma_2  R_c L / D + \gamma_3 \exp\left[-\beta G_0\right]
\end{equation}

Let us now turn to calculating $k_{\rm in}$, which we will do based on dimensional arguments. The dimension of $k_{\rm in}$ must be that of an area times an inverse time. The only physical quantities in our system that can be combined to give such a dimension basically tell us that $k_{\rm in}$ should be scaling as:
\begin{equation}
    k_{\rm in} = \gamma_1 D,
\end{equation}

$D$ again being the DNA diffusion coefficient and $\gamma_1$ a non-dimensional constant of order 1 that should only depend on the dimensionality of the system and on its geometry. Note that in practice the diffusion coefficient inside and outside the patch area might not be exactly the same. In fact, we would expect the diffusion of receptors within the patch area to be somewhat slower than outside of it, mostly because additional friction would be expected due to the interactions between unbound ligand-receptor pairs.\\
Note that given the previous expressions one should expect $k_{\rm in}$ to be the same for all of the systems studied here, whereas $k_{\rm out}$ should decrease for decreasing hybridisation energies $G_0$, i.e. for stronger bonds.

\paragraph*{{\normalfont \textbf{Acknowledgements}}}
{We thank Ramon van der Valk and Anne Schwabe for technical help and J\'er\'emie Capoulade for help with microscopy. D.V and S.A-U acknowledge the UK Materials and Molecular Modelling Hub for computational resources, which is partially funded by EPSRC (EP/P020194/1). CL, LL and DJK acknowledge support from the Netherlands Organization for Scientific Research (NWO/OCW), as part of the Gravitation Program: Frontiers of Nanoscience.

\bibliography{bibliography}

\begin{thebibliography}{44}
\expandafter\ifx\csname natexlab\endcsname\relax\def\natexlab#1{#1}\fi
\expandafter\ifx\csname bibnamefont\endcsname\relax
  \def\bibnamefont#1{#1}\fi
\expandafter\ifx\csname bibfnamefont\endcsname\relax
  \def\bibfnamefont#1{#1}\fi
\expandafter\ifx\csname citenamefont\endcsname\relax
  \def\citenamefont#1{#1}\fi
\expandafter\ifx\csname url\endcsname\relax
  \def\url#1{\texttt{#1}}\fi
\expandafter\ifx\csname urlprefix\endcsname\relax\def\urlprefix{URL }\fi
\providecommand{\bibinfo}[2]{#2}
\providecommand{\eprint}[2][]{\url{#2}}

\bibitem[{\citenamefont{Mammen et~al.}(1998)\citenamefont{Mammen, Choi, and
  Whitesides}}]{Mammen1998}
\bibinfo{author}{\bibfnamefont{M.}~\bibnamefont{Mammen}},
  \bibinfo{author}{\bibfnamefont{S.~K.} \bibnamefont{Choi}}, \bibnamefont{and}
  \bibinfo{author}{\bibfnamefont{G.~M.} \bibnamefont{Whitesides}},
  \bibinfo{journal}{Angewandte Chemie - International Edition}
  \textbf{\bibinfo{volume}{37}}, \bibinfo{pages}{2754} (\bibinfo{year}{1998}).

\bibitem[{\citenamefont{Huskens}(2006)}]{Huskens2006}
\bibinfo{author}{\bibfnamefont{J.}~\bibnamefont{Huskens}},
  \emph{\bibinfo{title}{{Multivalent interactions at interfaces}}}
  (\bibinfo{year}{2006}).

\bibitem[{\citenamefont{Satav et~al.}(2015)\citenamefont{Satav, Huskens, and
  Jonkheijm}}]{Satav2015}
\bibinfo{author}{\bibfnamefont{T.}~\bibnamefont{Satav}},
  \bibinfo{author}{\bibfnamefont{J.}~\bibnamefont{Huskens}}, \bibnamefont{and}
  \bibinfo{author}{\bibfnamefont{P.}~\bibnamefont{Jonkheijm}},
  \emph{\bibinfo{title}{{Effects of Variations in Ligand Density on Cell
  Signaling}}} (\bibinfo{year}{2015}).

\bibitem[{\citenamefont{Boudreau and Jones}(1999)}]{Boudreau1999}
\bibinfo{author}{\bibfnamefont{N.~J.} \bibnamefont{Boudreau}} \bibnamefont{and}
  \bibinfo{author}{\bibfnamefont{P.~L.} \bibnamefont{Jones}},
  \emph{\bibinfo{title}{{Extracellular matrix and integrin signalling: The
  shape of things to come}}} (\bibinfo{year}{1999}).

\bibitem[{\citenamefont{Fasting et~al.}(2012)\citenamefont{Fasting, Schalley,
  Weber, Seitz, Hecht, Koksch, Dernedde, Graf, Knapp, and Haag}}]{Fasting2012}
\bibinfo{author}{\bibfnamefont{C.}~\bibnamefont{Fasting}},
  \bibinfo{author}{\bibfnamefont{C.~A.} \bibnamefont{Schalley}},
  \bibinfo{author}{\bibfnamefont{M.}~\bibnamefont{Weber}},
  \bibinfo{author}{\bibfnamefont{O.}~\bibnamefont{Seitz}},
  \bibinfo{author}{\bibfnamefont{S.}~\bibnamefont{Hecht}},
  \bibinfo{author}{\bibfnamefont{B.}~\bibnamefont{Koksch}},
  \bibinfo{author}{\bibfnamefont{J.}~\bibnamefont{Dernedde}},
  \bibinfo{author}{\bibfnamefont{C.}~\bibnamefont{Graf}},
  \bibinfo{author}{\bibfnamefont{E.~W.} \bibnamefont{Knapp}}, \bibnamefont{and}
  \bibinfo{author}{\bibfnamefont{R.}~\bibnamefont{Haag}},
  \emph{\bibinfo{title}{{Multivalency as a chemical organization and action
  principle}}} (\bibinfo{year}{2012}).

\bibitem[{\citenamefont{Tian et~al.}(2020)\citenamefont{Tian,
  Angioletti-Uberti, and Battaglia}}]{Tian2020}
\bibinfo{author}{\bibfnamefont{X.}~\bibnamefont{Tian}},
  \bibinfo{author}{\bibfnamefont{S.}~\bibnamefont{Angioletti-Uberti}},
  \bibnamefont{and}
  \bibinfo{author}{\bibfnamefont{G.}~\bibnamefont{Battaglia}},
  \bibinfo{journal}{Science Advances} \textbf{\bibinfo{volume}{6}},
  \bibinfo{pages}{eaat0919} (\bibinfo{year}{2020}).

\bibitem[{\citenamefont{Liu et~al.}(2020)\citenamefont{Liu, Apriceno, Sipin,
  Scarpa, Rodriguez-Arco, Poma, Marchello, Battaglia, and
  Angioletti-Uberti}}]{Meng2020}
\bibinfo{author}{\bibfnamefont{M.}~\bibnamefont{Liu}},
  \bibinfo{author}{\bibfnamefont{A.}~\bibnamefont{Apriceno}},
  \bibinfo{author}{\bibfnamefont{M.}~\bibnamefont{Sipin}},
  \bibinfo{author}{\bibfnamefont{E.}~\bibnamefont{Scarpa}},
  \bibinfo{author}{\bibfnamefont{L.}~\bibnamefont{Rodriguez-Arco}},
  \bibinfo{author}{\bibfnamefont{A.}~\bibnamefont{Poma}},
  \bibinfo{author}{\bibfnamefont{G.}~\bibnamefont{Marchello}},
  \bibinfo{author}{\bibfnamefont{G.}~\bibnamefont{Battaglia}},
  \bibnamefont{and}
  \bibinfo{author}{\bibfnamefont{S.}~\bibnamefont{Angioletti-Uberti}},
  \bibinfo{journal}{Nature communications} \textbf{\bibinfo{volume}{11}},
  \bibinfo{pages}{1} (\bibinfo{year}{2020}).

\bibitem[{\citenamefont{Hauert and Bhatia}(2014)}]{Hauert2014}
\bibinfo{author}{\bibfnamefont{S.}~\bibnamefont{Hauert}} \bibnamefont{and}
  \bibinfo{author}{\bibfnamefont{S.~N.} \bibnamefont{Bhatia}},
  \emph{\bibinfo{title}{Mechanisms of cooperation in cancer nanomedicine:
  Towards systems nanotechnology}} (\bibinfo{year}{2014}).

\bibitem[{\citenamefont{Wang et~al.}(2020)\citenamefont{Wang, Min, Eghtesadi,
  Kane, and Chilkoti}}]{Wang2020}
\bibinfo{author}{\bibfnamefont{J.}~\bibnamefont{Wang}},
  \bibinfo{author}{\bibfnamefont{J.}~\bibnamefont{Min}},
  \bibinfo{author}{\bibfnamefont{S.~A.} \bibnamefont{Eghtesadi}},
  \bibinfo{author}{\bibfnamefont{R.~S.} \bibnamefont{Kane}}, \bibnamefont{and}
  \bibinfo{author}{\bibfnamefont{A.}~\bibnamefont{Chilkoti}},
  \bibinfo{journal}{ACS Nano} \textbf{\bibinfo{volume}{14}},
  \bibinfo{pages}{372} (\bibinfo{year}{2020}).

\bibitem[{\citenamefont{Zhang et~al.}(2020{\natexlab{a}})\citenamefont{Zhang,
  Wang, Gao, and Wu}}]{Zhang2020_2}
\bibinfo{author}{\bibfnamefont{J.}~\bibnamefont{Zhang}},
  \bibinfo{author}{\bibfnamefont{Z.}~\bibnamefont{Wang}},
  \bibinfo{author}{\bibfnamefont{Y.}~\bibnamefont{Gao}}, \bibnamefont{and}
  \bibinfo{author}{\bibfnamefont{Z.-S.} \bibnamefont{Wu}},
  \bibinfo{journal}{Cite This: ACS Appl. Bio Mater}
  \textbf{\bibinfo{volume}{2020}}, \bibinfo{pages}{4521}
  (\bibinfo{year}{2020}{\natexlab{a}}).

\bibitem[{\citenamefont{Zhang et~al.}(2019)\citenamefont{Zhang, Cheng, Cao,
  Zhang, Yuan, Wu, and Wang}}]{Zhang2019}
\bibinfo{author}{\bibfnamefont{Y.}~\bibnamefont{Zhang}},
  \bibinfo{author}{\bibfnamefont{M.}~\bibnamefont{Cheng}},
  \bibinfo{author}{\bibfnamefont{J.}~\bibnamefont{Cao}},
  \bibinfo{author}{\bibfnamefont{Y.}~\bibnamefont{Zhang}},
  \bibinfo{author}{\bibfnamefont{Z.}~\bibnamefont{Yuan}},
  \bibinfo{author}{\bibfnamefont{Q.}~\bibnamefont{Wu}}, \bibnamefont{and}
  \bibinfo{author}{\bibfnamefont{W.}~\bibnamefont{Wang}},
  \bibinfo{journal}{Nanoscale} \textbf{\bibinfo{volume}{11}},
  \bibinfo{pages}{5005} (\bibinfo{year}{2019}).

\bibitem[{\citenamefont{Li et~al.}(2005)\citenamefont{Li, Huang, and
  Peng}}]{Li2005}
\bibinfo{author}{\bibfnamefont{S.}~\bibnamefont{Li}},
  \bibinfo{author}{\bibfnamefont{S.}~\bibnamefont{Huang}}, \bibnamefont{and}
  \bibinfo{author}{\bibfnamefont{S.~B.} \bibnamefont{Peng}},
  \bibinfo{journal}{International Journal of Oncology}
  \textbf{\bibinfo{volume}{27}}, \bibinfo{pages}{1329} (\bibinfo{year}{2005}).

\bibitem[{\citenamefont{Akhtar et~al.}(2014)\citenamefont{Akhtar, Ahamed,
  Alhadlaq, Alrokayan, and Kumar}}]{Akhtar2014}
\bibinfo{author}{\bibfnamefont{M.~J.} \bibnamefont{Akhtar}},
  \bibinfo{author}{\bibfnamefont{M.}~\bibnamefont{Ahamed}},
  \bibinfo{author}{\bibfnamefont{H.~A.} \bibnamefont{Alhadlaq}},
  \bibinfo{author}{\bibfnamefont{S.~A.} \bibnamefont{Alrokayan}},
  \bibnamefont{and} \bibinfo{author}{\bibfnamefont{S.}~\bibnamefont{Kumar}},
  \emph{\bibinfo{title}{Targeted anticancer therapy: Overexpressed receptors
  and nanotechnology}} (\bibinfo{year}{2014}).

\bibitem[{\citenamefont{Koenig et~al.}(2021)\citenamefont{Koenig, Das, Liu,
  K{\"{u}}mmerer, Gohr, Jenster, Schiffelers, Tesfamariam, Uchima, Wuerth
  et~al.}}]{Koenig2021}
\bibinfo{author}{\bibfnamefont{P.-A.} \bibnamefont{Koenig}},
  \bibinfo{author}{\bibfnamefont{H.}~\bibnamefont{Das}},
  \bibinfo{author}{\bibfnamefont{H.}~\bibnamefont{Liu}},
  \bibinfo{author}{\bibfnamefont{B.~M.} \bibnamefont{K{\"{u}}mmerer}},
  \bibinfo{author}{\bibfnamefont{F.~N.} \bibnamefont{Gohr}},
  \bibinfo{author}{\bibfnamefont{L.-M.} \bibnamefont{Jenster}},
  \bibinfo{author}{\bibfnamefont{L.~D.~J.} \bibnamefont{Schiffelers}},
  \bibinfo{author}{\bibfnamefont{Y.~M.} \bibnamefont{Tesfamariam}},
  \bibinfo{author}{\bibfnamefont{M.}~\bibnamefont{Uchima}},
  \bibinfo{author}{\bibfnamefont{J.~D.} \bibnamefont{Wuerth}},
  \bibnamefont{et~al.}, \bibinfo{journal}{Science}
  \textbf{\bibinfo{volume}{371}}, \bibinfo{pages}{eabe6230}
  (\bibinfo{year}{2021}).

\bibitem[{\citenamefont{Doherty and McMahon}(2009)}]{Doherty2009}
\bibinfo{author}{\bibfnamefont{G.~J.} \bibnamefont{Doherty}} \bibnamefont{and}
  \bibinfo{author}{\bibfnamefont{H.~T.} \bibnamefont{McMahon}},
  \emph{\bibinfo{title}{{Mechanisms of endocytosis}}} (\bibinfo{year}{2009}).

\bibitem[{\citenamefont{Martinez-Veracoechea and
  Frenkel}(2011)}]{Martinez-Veracoechea2011}
\bibinfo{author}{\bibfnamefont{F.~J.} \bibnamefont{Martinez-Veracoechea}}
  \bibnamefont{and} \bibinfo{author}{\bibfnamefont{D.}~\bibnamefont{Frenkel}},
  \bibinfo{journal}{Proceedings of the National Academy of Sciences of the
  United States of America} \textbf{\bibinfo{volume}{108}},
  \bibinfo{pages}{10963} (\bibinfo{year}{2011}).

\bibitem[{\citenamefont{Duncan and Bevan}(2015)}]{Duncan2015}
\bibinfo{author}{\bibfnamefont{G.~A.} \bibnamefont{Duncan}} \bibnamefont{and}
  \bibinfo{author}{\bibfnamefont{M.~A.} \bibnamefont{Bevan}},
  \bibinfo{journal}{Nanoscale} \textbf{\bibinfo{volume}{7}}
  (\bibinfo{year}{2015}).

\bibitem[{\citenamefont{Wang and Dormidontova}(2012)}]{Wang2012}
\bibinfo{author}{\bibfnamefont{S.}~\bibnamefont{Wang}} \bibnamefont{and}
  \bibinfo{author}{\bibfnamefont{E.~E.} \bibnamefont{Dormidontova}},
  \bibinfo{journal}{Physical Review Letters} \textbf{\bibinfo{volume}{109}}
  (\bibinfo{year}{2012}).

\bibitem[{\citenamefont{Dubacheva et~al.}(2019)\citenamefont{Dubacheva, Curk,
  Frenkel, and Richter}}]{Dubacheva2019}
\bibinfo{author}{\bibfnamefont{G.~V.} \bibnamefont{Dubacheva}},
  \bibinfo{author}{\bibfnamefont{T.}~\bibnamefont{Curk}},
  \bibinfo{author}{\bibfnamefont{D.}~\bibnamefont{Frenkel}}, \bibnamefont{and}
  \bibinfo{author}{\bibfnamefont{R.~P.} \bibnamefont{Richter}},
  \bibinfo{journal}{Journal of the American Chemical Society}
  \textbf{\bibinfo{volume}{141}}, \bibinfo{pages}{2577} (\bibinfo{year}{2019}).

\bibitem[{\citenamefont{Lanfranco et~al.}(2019)\citenamefont{Lanfranco, Jana,
  Tunesi, Cicuta, Mognetti, {Di Michele}, and Bruylants}}]{Lanfranco2019}
\bibinfo{author}{\bibfnamefont{R.}~\bibnamefont{Lanfranco}},
  \bibinfo{author}{\bibfnamefont{P.~K.} \bibnamefont{Jana}},
  \bibinfo{author}{\bibfnamefont{L.}~\bibnamefont{Tunesi}},
  \bibinfo{author}{\bibfnamefont{P.}~\bibnamefont{Cicuta}},
  \bibinfo{author}{\bibfnamefont{B.~M.} \bibnamefont{Mognetti}},
  \bibinfo{author}{\bibfnamefont{L.}~\bibnamefont{{Di Michele}}},
  \bibnamefont{and}
  \bibinfo{author}{\bibfnamefont{G.}~\bibnamefont{Bruylants}},
  \bibinfo{journal}{Langmuir} \textbf{\bibinfo{volume}{35}},
  \bibinfo{pages}{2002} (\bibinfo{year}{2019}).

\bibitem[{\citenamefont{Di~Iorio et~al.}(2020)\citenamefont{Di~Iorio, Lu,
  Meulman, and Huskens}}]{DiIorio2020}
\bibinfo{author}{\bibfnamefont{D.}~\bibnamefont{Di~Iorio}},
  \bibinfo{author}{\bibfnamefont{Y.}~\bibnamefont{Lu}},
  \bibinfo{author}{\bibfnamefont{J.}~\bibnamefont{Meulman}}, \bibnamefont{and}
  \bibinfo{author}{\bibfnamefont{J.}~\bibnamefont{Huskens}},
  \bibinfo{journal}{Chemical Science} \textbf{\bibinfo{volume}{11}},
  \bibinfo{pages}{3307} (\bibinfo{year}{2020}).

\bibitem[{\citenamefont{Dubacheva et~al.}(2014)\citenamefont{Dubacheva, Curk,
  Mognetti, Auz{\'{e}}ly-Velty, Frenkel, and Richter}}]{Dubacheva2014}
\bibinfo{author}{\bibfnamefont{G.~V.} \bibnamefont{Dubacheva}},
  \bibinfo{author}{\bibfnamefont{T.}~\bibnamefont{Curk}},
  \bibinfo{author}{\bibfnamefont{B.~M.} \bibnamefont{Mognetti}},
  \bibinfo{author}{\bibfnamefont{R.}~\bibnamefont{Auz{\'{e}}ly-Velty}},
  \bibinfo{author}{\bibfnamefont{D.}~\bibnamefont{Frenkel}}, \bibnamefont{and}
  \bibinfo{author}{\bibfnamefont{R.~P.} \bibnamefont{Richter}},
  \bibinfo{journal}{Journal of the American Chemical Society}
  \textbf{\bibinfo{volume}{136}}, \bibinfo{pages}{1722} (\bibinfo{year}{2014}).

\bibitem[{\citenamefont{Dubacheva et~al.}(2015)\citenamefont{Dubacheva, Curk,
  Auz{\'e}ly-Velty, Frenkel, and Richter}}]{Dubacheva2015}
\bibinfo{author}{\bibfnamefont{G.~V.} \bibnamefont{Dubacheva}},
  \bibinfo{author}{\bibfnamefont{T.}~\bibnamefont{Curk}},
  \bibinfo{author}{\bibfnamefont{R.}~\bibnamefont{Auz{\'e}ly-Velty}},
  \bibinfo{author}{\bibfnamefont{D.}~\bibnamefont{Frenkel}}, \bibnamefont{and}
  \bibinfo{author}{\bibfnamefont{R.~P.} \bibnamefont{Richter}},
  \bibinfo{journal}{Proceedings of the National Academy of Sciences of the
  United States of America} \textbf{\bibinfo{volume}{112}},
  \bibinfo{pages}{5579} (\bibinfo{year}{2015}).

\bibitem[{\citenamefont{Scheepers et~al.}(2020)\citenamefont{Scheepers, van
  IJzendoorn, and Prins}}]{Scheepers2020}
\bibinfo{author}{\bibfnamefont{M.~R.~W.} \bibnamefont{Scheepers}},
  \bibinfo{author}{\bibfnamefont{L.~J.} \bibnamefont{van IJzendoorn}},
  \bibnamefont{and} \bibinfo{author}{\bibfnamefont{M.~W.~J.}
  \bibnamefont{Prins}}, \bibinfo{journal}{Proceedings of the National Academy
  of Sciences} \textbf{\bibinfo{volume}{10}}, \bibinfo{pages}{202003968}
  (\bibinfo{year}{2020}).

\bibitem[{\citenamefont{Overeem et~al.}(2020)\citenamefont{Overeem, Hamming,
  Grant, Di~Iorio, Tieke, Bertolino, Li, Vos, de~Vries, Woods
  et~al.}}]{Overeem}
\bibinfo{author}{\bibfnamefont{N.~J.} \bibnamefont{Overeem}},
  \bibinfo{author}{\bibfnamefont{P.~H.~E.} \bibnamefont{Hamming}},
  \bibinfo{author}{\bibfnamefont{O.~C.} \bibnamefont{Grant}},
  \bibinfo{author}{\bibfnamefont{D.}~\bibnamefont{Di~Iorio}},
  \bibinfo{author}{\bibfnamefont{M.}~\bibnamefont{Tieke}},
  \bibinfo{author}{\bibfnamefont{M.~C.} \bibnamefont{Bertolino}},
  \bibinfo{author}{\bibfnamefont{Z.}~\bibnamefont{Li}},
  \bibinfo{author}{\bibfnamefont{G.}~\bibnamefont{Vos}},
  \bibinfo{author}{\bibfnamefont{R.~P.} \bibnamefont{de~Vries}},
  \bibinfo{author}{\bibfnamefont{R.~J.} \bibnamefont{Woods}},
  \bibnamefont{et~al.}, \bibinfo{journal}{ACS Central Science}
  \textbf{\bibinfo{volume}{6}}, \bibinfo{pages}{2311} (\bibinfo{year}{2020}).

\bibitem[{\citenamefont{Rinaldin et~al.}(2019)\citenamefont{Rinaldin, Verweij,
  Chakraborty, and Kraft}}]{Rinaldin2019}
\bibinfo{author}{\bibfnamefont{M.}~\bibnamefont{Rinaldin}},
  \bibinfo{author}{\bibfnamefont{R.~W.} \bibnamefont{Verweij}},
  \bibinfo{author}{\bibfnamefont{I.}~\bibnamefont{Chakraborty}},
  \bibnamefont{and} \bibinfo{author}{\bibfnamefont{D.~J.} \bibnamefont{Kraft}},
  \bibinfo{journal}{Soft Matter} \textbf{\bibinfo{volume}{15}},
  \bibinfo{pages}{1345} (\bibinfo{year}{2019}).

\bibitem[{\citenamefont{Aloia}(1985)}]{membrane}
\bibinfo{author}{\bibfnamefont{R.~C.} \bibnamefont{Aloia}},
  \emph{\bibinfo{title}{{Membrane Fluidity in Biology: Cellular Aspects}}}
  (\bibinfo{year}{1985}).

\bibitem[{\citenamefont{Zhang et~al.}(2020{\natexlab{b}})\citenamefont{Zhang,
  Li, Dong, and Han}}]{Zhang2020}
\bibinfo{author}{\bibfnamefont{Y.}~\bibnamefont{Zhang}},
  \bibinfo{author}{\bibfnamefont{Q.}~\bibnamefont{Li}},
  \bibinfo{author}{\bibfnamefont{M.}~\bibnamefont{Dong}}, \bibnamefont{and}
  \bibinfo{author}{\bibfnamefont{X.}~\bibnamefont{Han}},
  \bibinfo{journal}{Colloids and Surfaces B: Biointerfaces}
  \textbf{\bibinfo{volume}{196}}, \bibinfo{pages}{111353}
  (\bibinfo{year}{2020}{\natexlab{b}}).

\bibitem[{\citenamefont{Chakraborty et~al.}(2020)\citenamefont{Chakraborty,
  Doktorova, Molugu, Heberle, Scott, Dzikovski, Nagao, Stingaciu, Standaert,
  Barrera et~al.}}]{Chakraborty2020}
\bibinfo{author}{\bibfnamefont{S.}~\bibnamefont{Chakraborty}},
  \bibinfo{author}{\bibfnamefont{M.}~\bibnamefont{Doktorova}},
  \bibinfo{author}{\bibfnamefont{T.~R.} \bibnamefont{Molugu}},
  \bibinfo{author}{\bibfnamefont{F.~A.} \bibnamefont{Heberle}},
  \bibinfo{author}{\bibfnamefont{H.~L.} \bibnamefont{Scott}},
  \bibinfo{author}{\bibfnamefont{B.}~\bibnamefont{Dzikovski}},
  \bibinfo{author}{\bibfnamefont{M.}~\bibnamefont{Nagao}},
  \bibinfo{author}{\bibfnamefont{L.~R.} \bibnamefont{Stingaciu}},
  \bibinfo{author}{\bibfnamefont{R.~F.} \bibnamefont{Standaert}},
  \bibinfo{author}{\bibfnamefont{F.~N.} \bibnamefont{Barrera}},
  \bibnamefont{et~al.}, \bibinfo{journal}{Proceedings of the National Academy
  of Sciences of the United States of America} \textbf{\bibinfo{volume}{117}},
  \bibinfo{pages}{21896} (\bibinfo{year}{2020}).

\bibitem[{\citenamefont{Delcanale et~al.}(2018)\citenamefont{Delcanale,
  Miret-Ontiveros, Arista-Romero, Pujals, and Albertazzi}}]{Delcanale2018}
\bibinfo{author}{\bibfnamefont{P.}~\bibnamefont{Delcanale}},
  \bibinfo{author}{\bibfnamefont{B.}~\bibnamefont{Miret-Ontiveros}},
  \bibinfo{author}{\bibfnamefont{M.}~\bibnamefont{Arista-Romero}},
  \bibinfo{author}{\bibfnamefont{S.}~\bibnamefont{Pujals}}, \bibnamefont{and}
  \bibinfo{author}{\bibfnamefont{L.}~\bibnamefont{Albertazzi}},
  \bibinfo{journal}{ACS Nano} \textbf{\bibinfo{volume}{12}},
  \bibinfo{pages}{7629} (\bibinfo{year}{2018}).

\bibitem[{\citenamefont{Varilly et~al.}(2012)\citenamefont{Varilly,
  Angioletti-Uberti, Mognetti, and Frenkel}}]{Varilly2012}
\bibinfo{author}{\bibfnamefont{P.}~\bibnamefont{Varilly}},
  \bibinfo{author}{\bibfnamefont{S.}~\bibnamefont{Angioletti-Uberti}},
  \bibinfo{author}{\bibfnamefont{B.~M.} \bibnamefont{Mognetti}},
  \bibnamefont{and} \bibinfo{author}{\bibfnamefont{D.}~\bibnamefont{Frenkel}},
  \bibinfo{journal}{The Journal of chemical physics}
  \textbf{\bibinfo{volume}{137}}, \bibinfo{pages}{094108}
  (\bibinfo{year}{2012}).

\bibitem[{\citenamefont{Angioletti-Uberti
  et~al.}(2014)\citenamefont{Angioletti-Uberti, Varilly, Mognetti, and
  Frenkel}}]{Angioletti2014}
\bibinfo{author}{\bibfnamefont{S.}~\bibnamefont{Angioletti-Uberti}},
  \bibinfo{author}{\bibfnamefont{P.}~\bibnamefont{Varilly}},
  \bibinfo{author}{\bibfnamefont{B.~M.} \bibnamefont{Mognetti}},
  \bibnamefont{and} \bibinfo{author}{\bibfnamefont{D.}~\bibnamefont{Frenkel}},
  \bibinfo{journal}{Physical review letters} \textbf{\bibinfo{volume}{113}},
  \bibinfo{pages}{128303} (\bibinfo{year}{2014}).

\bibitem[{\citenamefont{Mognetti et~al.}(2019)\citenamefont{Mognetti, Cicuta,
  and Di~Michele}}]{mognetti2019}
\bibinfo{author}{\bibfnamefont{B.~M.} \bibnamefont{Mognetti}},
  \bibinfo{author}{\bibfnamefont{P.}~\bibnamefont{Cicuta}}, \bibnamefont{and}
  \bibinfo{author}{\bibfnamefont{L.}~\bibnamefont{Di~Michele}},
  \bibinfo{journal}{Reports on Progress in Physics}
  \textbf{\bibinfo{volume}{82}}, \bibinfo{pages}{116601}
  (\bibinfo{year}{2019}).

\bibitem[{\citenamefont{SantaLucia}(1998)}]{Santalucia1998}
\bibinfo{author}{\bibfnamefont{J.}~\bibnamefont{SantaLucia}},
  \bibinfo{journal}{Proceedings of the National Academy of Sciences}
  \textbf{\bibinfo{volume}{95}}, \bibinfo{pages}{1460} (\bibinfo{year}{1998}).

\bibitem[{\citenamefont{Azizian}(2004)}]{aziz}
\bibinfo{author}{\bibfnamefont{S.}~\bibnamefont{Azizian}},
  \bibinfo{journal}{Journal of colloid and Interface Science}
  \textbf{\bibinfo{volume}{276}}, \bibinfo{pages}{47} (\bibinfo{year}{2004}).

\bibitem[{\citenamefont{Curk et~al.}(2017)\citenamefont{Curk, Dobnikar, and
  Frenkel}}]{Curk2017}
\bibinfo{author}{\bibfnamefont{T.}~\bibnamefont{Curk}},
  \bibinfo{author}{\bibfnamefont{J.}~\bibnamefont{Dobnikar}}, \bibnamefont{and}
  \bibinfo{author}{\bibfnamefont{D.}~\bibnamefont{Frenkel}},
  \bibinfo{journal}{Proceedings of the National Academy of Sciences of the
  United States of America} \textbf{\bibinfo{volume}{114}},
  \bibinfo{pages}{7210} (\bibinfo{year}{2017}).

\bibitem[{\citenamefont{Bray et~al.}(1998)\citenamefont{Bray, Levin, and
  Morton-Firth}}]{Bray1998}
\bibinfo{author}{\bibfnamefont{D.}~\bibnamefont{Bray}},
  \bibinfo{author}{\bibfnamefont{M.~D.} \bibnamefont{Levin}}, \bibnamefont{and}
  \bibinfo{author}{\bibfnamefont{C.~J.} \bibnamefont{Morton-Firth}},
  \bibinfo{journal}{Nature} \textbf{\bibinfo{volume}{393}}, \bibinfo{pages}{85}
  (\bibinfo{year}{1998}).

\bibitem[{\citenamefont{Stefanick et~al.}(2019)\citenamefont{Stefanick,
  Omstead, Kiziltepe, and Bilgicer}}]{Stefanick2019}
\bibinfo{author}{\bibfnamefont{J.~F.} \bibnamefont{Stefanick}},
  \bibinfo{author}{\bibfnamefont{D.~T.} \bibnamefont{Omstead}},
  \bibinfo{author}{\bibfnamefont{T.}~\bibnamefont{Kiziltepe}},
  \bibnamefont{and} \bibinfo{author}{\bibfnamefont{B.}~\bibnamefont{Bilgicer}},
  \bibinfo{journal}{Nanoscale} \textbf{\bibinfo{volume}{11}},
  \bibinfo{pages}{4414} (\bibinfo{year}{2019}).

\bibitem[{\citenamefont{Van Der~Meulen and Leunissen}(2013)}]{VanDerMeulen2013}
\bibinfo{author}{\bibfnamefont{S.~A.} \bibnamefont{Van Der~Meulen}}
  \bibnamefont{and} \bibinfo{author}{\bibfnamefont{M.~E.}
  \bibnamefont{Leunissen}}, \bibinfo{journal}{Journal of the American Chemical
  Society} \textbf{\bibinfo{volume}{135}}, \bibinfo{pages}{15129}
  (\bibinfo{year}{2013}).

\bibitem[{\citenamefont{Van~Ginkel et~al.}(2018)\citenamefont{Van~Ginkel,
  Filius, Szczepaniak, Tulinski, Meyer, and Joo}}]{VanGinkel2018}
\bibinfo{author}{\bibfnamefont{J.}~\bibnamefont{Van~Ginkel}},
  \bibinfo{author}{\bibfnamefont{M.}~\bibnamefont{Filius}},
  \bibinfo{author}{\bibfnamefont{M.}~\bibnamefont{Szczepaniak}},
  \bibinfo{author}{\bibfnamefont{P.}~\bibnamefont{Tulinski}},
  \bibinfo{author}{\bibfnamefont{A.~S.} \bibnamefont{Meyer}}, \bibnamefont{and}
  \bibinfo{author}{\bibfnamefont{C.}~\bibnamefont{Joo}},
  \bibinfo{journal}{Proceedings of the National Academy of Sciences of the
  United States of America} \textbf{\bibinfo{volume}{115}},
  \bibinfo{pages}{3338} (\bibinfo{year}{2018}).

\bibitem[{\citenamefont{Preibisch et~al.}(2010)\citenamefont{Preibisch,
  Saalfeld, Schindelin, and Tomancak}}]{Preibisch2010}
\bibinfo{author}{\bibfnamefont{S.}~\bibnamefont{Preibisch}},
  \bibinfo{author}{\bibfnamefont{S.}~\bibnamefont{Saalfeld}},
  \bibinfo{author}{\bibfnamefont{J.}~\bibnamefont{Schindelin}},
  \bibnamefont{and} \bibinfo{author}{\bibfnamefont{P.}~\bibnamefont{Tomancak}},
  \emph{\bibinfo{title}{{Software for bead-based registration of selective
  plane illumination microscopy data}}} (\bibinfo{year}{2010}).

\bibitem[{\citenamefont{Angioletti-Uberti
  et~al.}(2013)\citenamefont{Angioletti-Uberti, Varilly, Mognetti, Tkachenko,
  and Frenkel}}]{Angioletti2013}
\bibinfo{author}{\bibfnamefont{S.}~\bibnamefont{Angioletti-Uberti}},
  \bibinfo{author}{\bibfnamefont{P.}~\bibnamefont{Varilly}},
  \bibinfo{author}{\bibfnamefont{B.~M.} \bibnamefont{Mognetti}},
  \bibinfo{author}{\bibfnamefont{A.~V.} \bibnamefont{Tkachenko}},
  \bibnamefont{and} \bibinfo{author}{\bibfnamefont{D.}~\bibnamefont{Frenkel}},
  \emph{\bibinfo{title}{Communication: A simple analytical formula for the free
  energy of ligand--receptor-mediated interactions}} (\bibinfo{year}{2013}).

\bibitem[{\citenamefont{Angioletti-Uberti}(2017)}]{Angioletti2017}
\bibinfo{author}{\bibfnamefont{S.}~\bibnamefont{Angioletti-Uberti}},
  \bibinfo{journal}{Physical Review Letters} \textbf{\bibinfo{volume}{118}},
  \bibinfo{pages}{068001} (\bibinfo{year}{2017}).

\bibitem[{\citenamefont{Sanstead and Tokmakoff}(2018)}]{unbinding}
\bibinfo{author}{\bibfnamefont{P.~J.} \bibnamefont{Sanstead}} \bibnamefont{and}
  \bibinfo{author}{\bibfnamefont{A.}~\bibnamefont{Tokmakoff}},
  \bibinfo{journal}{The Journal of Physical Chemistry B}
  \textbf{\bibinfo{volume}{122}}, \bibinfo{pages}{3088} (\bibinfo{year}{2018}).

\end{thebibliography}

\end{document}